\documentclass[reprint,superscriptaddress,amsmath,amssymb,aps,pra,floatfix]{revtex4-1}

\usepackage[utf8]{inputenc}
\usepackage{graphicx}
\usepackage{dcolumn}
\usepackage{bm}
\usepackage{hyperref}

\usepackage{hypernat}

\usepackage[separate-uncertainty=true]{siunitx} 

\def\tol#1#2#3{\hbox{\rule{0pt}{15pt}${#1}^{+{#2}}_{-{#3}}$}}

\def\@fnsymbol#1{\ensuremath{\ifcase#1\or *\or \dagger\or \ddagger\or
   \mathsection\or \mathparagraph\or \|\or **\or \dagger\dagger
   \or \ddagger\ddagger \else\@ctrerr\fi}}

\DeclareRobustCommand*{\changesfromto}[2]{
{#2}}

\begin{document}

\title{Overcoming the limitations of the MARTINI force field in Molecular Dynamics simulations of polysaccharides}

\author{Philipp S. Schmalhorst}
\affiliation{Institute of Science and Technology Austria, Am Campus 1, 3400 Klosterneuburg, Austria}

\author{Felix Deluweit}

\author{Roger Scherrers}
\affiliation{Wyatt Technology Europe, Hochstra{\ss}e 18, 56307 Dernbach, Germany}

\author{Carl-Philipp Heisenberg}

\author{Mateusz Sikora}
\email{msikora@ist.ac.at}
\affiliation{Institute of Science and Technology Austria, Am Campus 1, 3400 Klosterneuburg, Austria}

\date{\today}

\begin{abstract}
Polysaccharides (carbohydrates) are key regulators of a large number of cell biological processes.
 However, precise biochemical or genetic manipulation of these often complex structures is laborious and hampers experimental structure-function studies.
 Molecular Dynamics (MD) simulations provide a valuable alternative tool to generate and test hypotheses on saccharide function.
 Yet, \changesfromto{most currently used MD force fields overstate non-bonded interactions between polysaccharide molecules, strongly limiting the usability of those simulations.}{currently used MD force fields often overestimate the aggregation propensity of polysaccharides, affecting the usability of those simulations.}
 Here we tested MARTINI, a popular coarse-grained (CG) force field for biological macromolecules, for its ability to accurately represent molecular forces between saccharides. To this end, we calculated a thermodynamic solution property, the second virial coefficient of the osmotic pressure ($B_{22}$). Comparison with light scattering experiments revealed \changesfromto{that MARTINI strongly overestimates non-bonded interactions for a prototypical polysaccharide, leading to unrealistic aggregation.}{a non-physical aggregation of a prototypical polysaccharide in MARTINI, pointing at an imbalance of the non-bonded solute-solute, solute-water, and water-water interactions.}
 This finding also applies to smaller oligosaccharides which were all found to aggregate in simulations even at moderate concentrations, well below their solubility limit.
 Finally, we explored the influence of the Lennard-Jones (LJ) interaction between saccharide molecules and propose a simple scaling of the LJ interaction strength that makes MARTINI more reliable for the simulation of saccharides.
\end{abstract}

\maketitle

\section{INTRODUCTION}
\paragraph*{Polysaccharides}
Polysaccharides are sugar polymers found in various biological contexts, {\it e.g.} glycoproteins, proteoglycans, bacterial lipopolysaccharides, and cell walls of plants and fungi, functioning as biomolecular interaction modulators\cite{Johnson2005,Langer2012,Moremen2012}, structural elements and energy storage. Their structural diversity ranges from simple linear homo-polymers, like cellulose, to cyclic or branched structures composed of diverse sugars connected via glycosidic linkages\cite{glycobiology} (fig.~\ref{fgr:mapping}). This complexity and the often encountered microheterogeneity, {\it i.e.} the simultaneous occurrence of length and structure variants of a given polysaccharide, hamper experimental approaches to study the roles of polysaccharides in biological processes. This gap can be filled by molecular dynamics (MD) simulations if the representation of polysaccharides in the respective model accurately captures their biophysical properties. Many carbohydrate-specific force fields have been developed (see \citeauthor{Foley2012}\cite{Foley2012} for a review) and applied to study the behavior of saccharides in solutions as well as their interactions with other biomolecules\cite{Feng2015,Xiong2015,Sauter2017}. In this regard, the non-bonded interactions are of particular importance as they determine the magnitude of intermolecular forces. In MD force fields they are typically represented by Lennard-Jones (LJ) and Coulomb potentials for van der Waals and electrostatic interactions, respectively.

\paragraph*{The MARTINI coarse-grained force field}
The goal of MD simulations is to extract biophysical properties from a system that is incrementally evolved over time.
This essentially means averaging over stochastic events, which requires system size and simulation length to be sufficiently large.
Simulations at atomistic level (all-atom, AA) require femtosecond time increments to model bond stretching and so they are typically limited to system sizes on the order of $10^5$ atoms and time scales below 1 $\mu$s.
This is insufficient for many processes involving polysaccharides in terms of both size and time.
A way to overcome these limitations is the use of coarse-grained (CG) force fields which sacrifice atomic-resolution detail to reduce computational cost.
The usual strategy involves replacing groups of atoms with larger pseudo-atoms (beads) that retain averaged properties of the underlying atomistic system.
This reduces the number of particles in the system and smooths the free energy landscape, increasing the effective time span covered by the simulation.

MARTINI\cite{Marrink2007} is a CG force field that maps groups of four neighbor "heavy" atoms (C, O, N, P, S) onto pre-defined beads.
This rather modest level of coarse-graining retains much structural detail while offering substantial speed-up compared to all-atom simulations.
Although originally devised for lipid systems, MARTINI has been extended to proteins\cite{monticelli_martini_2008}, polysaccharides\cite{Lopez2009} and nucleic acids\cite{Uusitalo2015}.
When transferring an atomistic system to MARTINI, parameters for bonded interactions between the beads have to be found empirically, {\it e.g.} by comparison to an AA simulation, while non-bonded interactions are predetermined by bead type and charge.
MARTINI beads are not endowed with partial charges of the component heavy atoms, so non-bonded interactions of uncharged molecules are determined by LJ interactions only.
The LJ potential $V_{LJ}$ between two beads $i$ and $j$ at a distance $r_{ij}$ takes the following shape:

\begin{equation}
V_{LJ}(r_{ij})=4\epsilon_{ij}\left[\left(\frac{\sigma_{ij}}{r_{ij}}\right)^{12}-\left(\frac{\sigma_{ij}}{r_{ij}}\right)^{6}\right]
\label{eqn:LJ_potential}
\end{equation}

The finite distance at which the potential reaches zero, $\sigma_{ij}$, and the depth of the potential well, $\epsilon_{ij}$, vary with the bead type and have been fit to reproduce partition coefficients of small reference molecules in polar/apolar solvent systems\cite{Marrink2007}.

\paragraph*{The problem of non-bonded interactions}
This common strategy of simply extrapolating from small molecule interaction energies to the macromolecules of interest bears the risk of misrepresenting the strength of interactions between macromolecules due to multiplication of \changesfromto{none}{small} systematic errors in the parametrization.
Indeed, it has been found in recent years that \changesfromto{modern AA force fields overestimate the intermolecular interaction strength of proteins}{AA simulations tend to overestimate the aggregation propensity of proteins} \cite{Petrov2014,Best_2014_aggregation,Jao_2015_aggregation} and polysaccharides\cite{Sauter2015}.
The widely accepted notion is that \changesfromto{none}{this is due to an imbalance of solute-solute, solute-water, and water-water interactions which are determined by the choice of the water model and the solute-solute interaction potentials defined in the force field. 
To alleviate this imbalance}, it has been proposed to include experimentally addressable solution properties in the parametrization process, such as the osmotic pressure\cite{Luo2010,Yoo2012,Yoo2016}, Kirkwood-Buff integrals\cite{Ploetz2011,Karunaweera2012} or the osmotic coefficient\cite{Miller2016_aa,Miller2017}.
All these parameters relate to the second virial coefficient, $B_{22}$, of the osmotic pressure, $\Pi$, which describes the deviation from ideal behavior of a solution with solute molar concentration $c$\cite{McMillan1945}:

\begin{equation}
\Pi(T,c)=RT(c+B_{22}c^2+B_{23}c^3+\dots)
\label{eqn:osmoticP_mole}
\end{equation}

Here, $T$ denotes temperature, $R$ is the gas constant and $B_{ij}$ are coefficients of the virial expansion, with $i=2$ referring to the solute in a binary mixture and $j$ enumerating consecutive coefficients.
$B_{22}>0$ indicates net attraction and $B_{22}<0$ repulsion between molecules, whereas the magnitude corresponds to the aggregation propensity.
Experimentally, $B_{22}$ can be determined {\it e.g.} through direct measurements of $\Pi$\cite{Stigter1960}, self-interaction chromatography\cite{tessier_self-interaction_2002}, or diffraction experiments\cite{George1997}.
These methods are complemented by an established theoretical groundwork\cite{McMillan1945} that allows for calculation of $B_{22}$ from MD simulations, making it a powerful tool for the refinement of intermolecular interactions.
Such refinement has been recently performed for several force fields\cite{Blanco_B22,Likos_B22,Hummer_B22,Grunberger_B22,Rubio_B22}, including a study from the Elcock group\cite{Stark2013}, who found \changesfromto{MARTINI protein-protein interaction strengths to be dramatically overestimated, resulting in}{} an abnormally strong aggregation propensity \changesfromto{none}{of proteins in MARTINI} that could be remedied by reduction of $\epsilon_{ij}$ of solute-solute LJ interactions.
We hypothesize that a similar problem exists for polysaccharides in MARTINI, and furthermore, due to a cumulative effect of many interacting atoms, this spurious aggregation propensity is expected to grow with the size of interacting molecules. 
Should this be the case, it would pose a threat that unrealistic aggregation behavior could influence outcomes of studies based on the present MARTINI model for saccharides (e.g.\cite{LPS,lopez_cellulose_2015,Kapla2016}).
In this work we compute $B_{22}$ from MARTINI simulations of five different saccharides (1 to 11 residues) and compare these to experimental values. To this end we utilize light scattering experiments to determine $B_{22}$ of a complex branched polysaccharide prototypical for protein glycosylation.
We demonstrate that MARTINI considerably overestimates the aggregation behavior of even small saccharides and show that uniform scaling of the LJ parameter $\epsilon_{ij}$ is sufficient to facilitate more realistic outcomes of MARTINI simulations.

\section{METHODS}

\subsection{Computational methods}

\subsubsection{MD simulations}
All simulations in this work were performed using GROMACS\cite{Abraham201519} 5.0.4 for enhanced sampling simulations and 5.1 otherwise. 

\paragraph*{All atom simulations}
All atom simulations of the carbohydrates glucose, sucrose, $\alpha$-cyclodextrin ($\alpha$-CD), $\beta$-cyclodextrin ($\beta$-CD), 
and A2 glycan were conducted using the \changesfromto{none}{GLYCAM06\cite{glycam06}} force field and initial structures and carbohydrate-specific force field parameters were obtained from the GLYCAM06j-1 carbohydrate builder suite\cite{glycam06} except for CDs, which were manually prepared based on the corresponding amylohexa- and -heptaose structures.
Topologies were converted to GROMACS format using ACPYPE\cite{SousadaSilva2012}.
For each of the saccharides, the initial structure was placed in a cubic box large enough to prevent self-interactions, subjected to energy minimization and solvated with SPC water\cite{SPC}.
Subsequently, \changesfromto{sodium and chloride ions were added}{a number of water beads was replaced with sodium and chloride ions in order} to neutralize any charges and to obtain a salt concentration of approximately physiological 100~mM NaCl.
Finally, the system was equilibrated in the canonical ensemble, ({\it i.e.} with constant volume and temperature) for 250~ps.
Production runs were performed in the isothermal-isobaric ensemble, with Nos\'e-Hoover\cite{nose} thermostat and Parrinello-Rahman\cite{parinello} barostat, with long range electrostatic interactions calculated by the Particle-Mesh Ewald method and short-range electrostatics and van der Waals interactions cut off at 1.0~nm, as recommended for the GLYCAM06 force field\cite{glycam06}. All simulations were performed at a temperature of 300~K using an integration step of 2~fs. Trajectories of 150~ns were used for mapping of bonded interactions. 

\begin{figure*}[ht!]
  \includegraphics[width=0.6\textwidth]{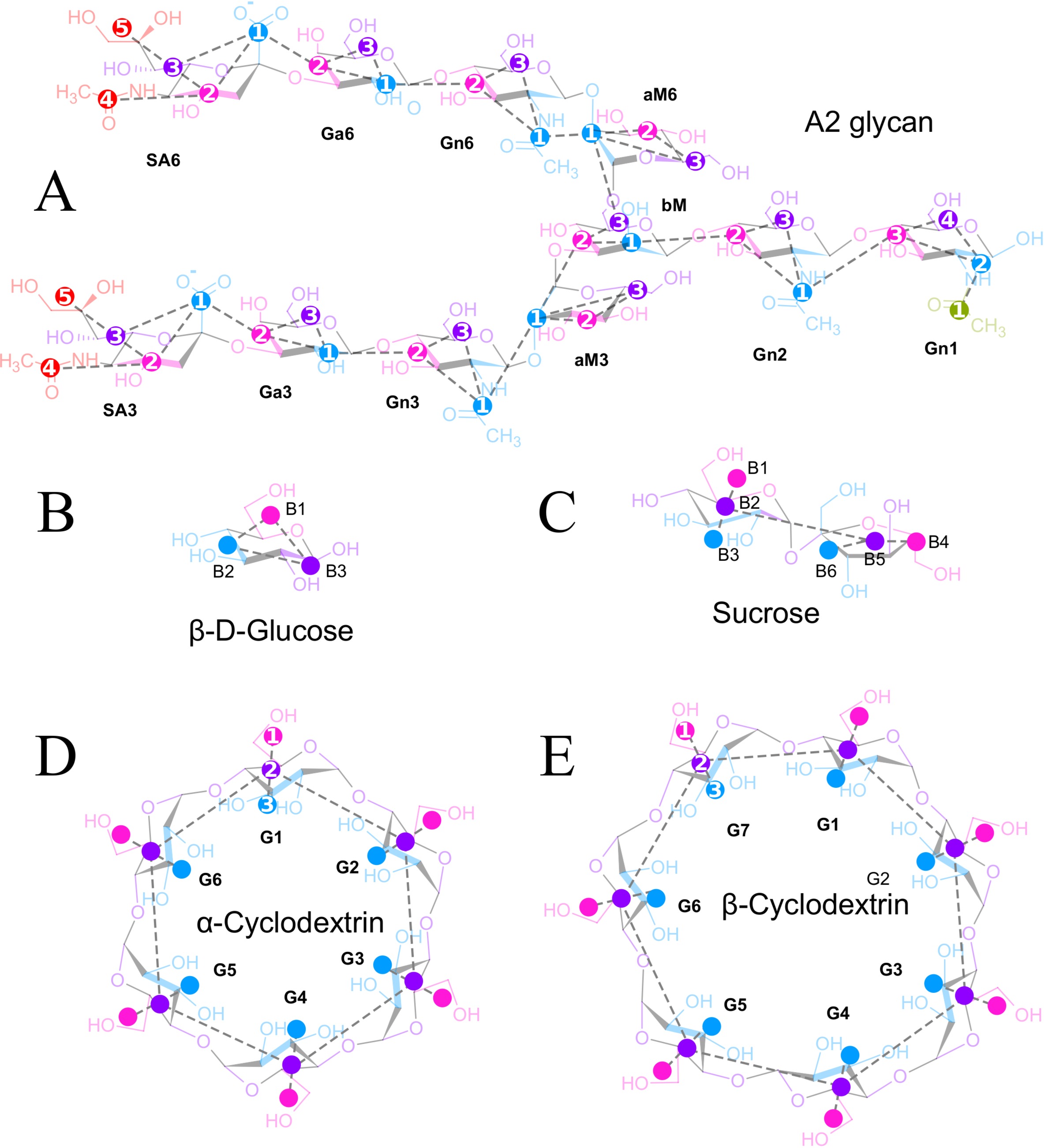}
  \caption{MARTINI mapping schemes for \changesfromto{none}{the five} saccharides studied in this manuscript. Semitransparent atomistic  structures were overlaid with coarse-grained representations (filled circles). Corresponding colors indicate how atoms were assigned to beads. Where necessary, bead numbers are given in white. Dashed lines represent connectivity between coarse-grained beads.}
  \label{fgr:mapping}
  \end{figure*}

\paragraph*{Parametrization of saccharides in the MARTINI force field}
We generally followed the proposed parametrization scheme for monosaccharides and linear oligo- and polysaccharides\cite{Lopez2009}. A monosaccharide unit is therein represented by three ,,polar'' beads which can be of type $\mathrm{P_1}$, $\mathrm{P_2}$, $\mathrm{P_3}$, or $\mathrm{P_4}$. The differences in bead polarity, denoted by the subscripts, are reflected in different strengths of the LJ interaction parameter $\epsilon$. The full interaction matrix is found in the original MARTINI force field publication\cite{Marrink2007}. For glucose and sucrose we adopted the published parameter set\cite{Lopez2009}. Glucose therein consists of a triangle of three beads, two more polar $\mathrm{P_4}$ beads and one less polar $\mathrm{P_1}$ bead (see fig.~\ref{fgr:mapping}B and Tables~\ref{tbl:bead_types} and \ref{tbl:params} for topology and corresponding bead types). The two sugar units in sucrose are represented by a different topology. A central bead connects to two outer beads that are not inter-connected. The monosaccharide units are linked through a bond between their central beads. The employed bead types are $\mathrm{P_1}$, $\mathrm{P_2}$, and $\mathrm{P_4}$ (fig.~\ref{fgr:mapping}C).
For the cyclic oligosaccharides $\alpha$- and $\beta$-CD (fig.~\ref{fgr:mapping}D and E) the parameters\cite{Lopez2009} for maltose ($\alpha$-glucosyl-(1$\rightarrow$4)-$\alpha$-glucose) and previous work on $\beta$-CD \cite{Lopez2013_CD_MARTINI} served as template for a mapping procedure using AA simulations as a reference.

To this end, we iterated through short 50~ns CG simulations (described below), extracted bond and angle distributions and made small adjustments to bond and angle (and dihedral, see below) force field parameters until satisfactory agreement with reference distributions obtained from {\it a posteriori} coarse grained AA simulations had been reached\cite{Lopez2009}.
The same procedure was used to parametrize A2 glycan (fig.~\ref{fgr:mapping}A), for which the mapping rules put forward for oligosaccharides\cite{Lopez2009} together with the parameters of the glycolipid GM1\cite{Lopez2013_glycolipids} provided the starting point. Herein, the additional bead types $\mathrm{SP_1}$, $\mathrm{P_5}$, $\mathrm{N_a}$, $\mathrm{N_{da}}$, and $\mathrm{Q_a}$ were used.
Unfortunately, the introduction of dihedral potentials for branched A2 glycan caused instability of simulations unless a time step of 5~fs instead of 30~fs was used, as has been noted before\cite{Lopez2013_glycolipids}.
The dihedral potentials were therefore used for sucrose, but disabled in A2 glycan parameters.
Nevertheless the glycosidic dihedrals, being most instructive for the conformation of A2 glycan, in most cases show qualitatively similar behavior in MARTINI and AA simulations (see the supplementary files for comparison of AA to CG distributions with dihedral potentials switched on and off).
The complete parameter set for the bonded interactions is listed in table~\ref{tbl:params}.

\paragraph*{Coarse-grained simulations} For all coarse-grained simulations the MARTINI force field version 2.2 was used, with modifications as described in the Results section. 
Briefly, the {\it martinize} script (v2.4), obtained from the MARTINI website\cite{martinize} and supplemented with mapping schemes and bonded parameters of saccharides studied in this work, was used to convert atomistic structures into coarse-grained ones and to generate the necessary topology files.
Similarly to AA simulations, structures were energy-minimized, solvated with MARTINI water containing 100~mM NaCl and 10~mM $\mathrm{CaCl_2}$ and equilibrated in the canonical ensemble for a total of 1~ns. \changesfromto{none}{Additional tests showed that the inclusion of calcium ions, not thoroughly tested in MARTINI, had no influence on the aggregating properties of saccharides (see fig.~\ref{SI:ions_WF}B).}
Production runs were performed in the isothermal-isobaric ensemble using velocity-rescale\cite{vrescale} and Parrinello-Rahman coupling schemes to keep temperature and pressure constant. For all MARTINI simulations a time step of 30~fs was used. To allow for GPU-accelerated simulations, the recently described Verlet neighbor search algorithm was used\cite{deJong20161}. Briefly, both LJ and Coulomb potentials were cut off beyond 1.1~nm. The LJ potential was in addition shifted to zero at the cutoff distance. For Coulomb interactions, the reaction-field potential was used. Details of the parameters were kept according to the recommended mdp files (http://cgmartini.nl).
Water is modeled in MARTINI as uncharged $\mathrm{P_4}$ beads with each bead representing four water molecules\cite{Marrink2007}. Special ''antifreeze'' beads have been introduced to avoid the freezing of water at temperatures around 300~K\cite{Marrink2007}. Unless otherwise stated 10\% of the water particles (W) are replaced by antifreeze particles (WF) \changesfromto{none}{(referred to as antifreeze water)} in the course of this study.

\subsubsection{Computation of \texorpdfstring{$B_{22}$}{B22}}
Two methods were employed to calculate $B_{22}$ from MD simulations, based on either the cumulative solute-solute radial distribution function (RDF), or a reconstruction of the potential of mean force (PMF) between two solute molecules over their separation distance.

\paragraph*{Cumulative RDF method}
McMillan and Mayer\cite{McMillan1945} derived an expression for $B_{22}$ under the assumption that the total solute potential energy can be approximated as the sum of pairwise solute-solute interactions, {\it i.e.} the potential of mean force (PMF) $W_2(r)$ between two solute particles at distance $r$:

\begin{equation}
B_{22}=-2\pi N_A \int_{0}^{\infty}[\exp(-\frac{W_2(r)}{RT})-1]r^2dr
\label{eqn:B22_pmf}
\end{equation}

where $N_A$ denotes Avogadro's constant. 
For weakly interacting solute particles, $B_{22}$ can be calculated efficiently by means of the radial distribution function (RDF), $g(r)$, which in thermodynamic equilibrium is approximately Boltzmann-distributed according to $W_2(r)$:

\begin{equation}
g(r)\approx \exp\left({-\frac{W_2(r)}{RT}}\right) 
\label{eqn:Boltzmann}
\end{equation}

Inserting \eqref{eqn:Boltzmann} into \eqref{eqn:B22fromPMF} gives:  

\begin{equation}\label{eqn:B22fromRDF}
B_{22}(r^\prime) 
=-2\pi N_A\int_{0}^{r^\prime}[g(r)-1]r^2dr
\end{equation}

Equation \eqref{eqn:B22fromRDF} can be simplified by making use of the definition  $g(r)$ as the solute molecule number increment d$N$ in the spherical shell d$V$ at distance $r$, normalized by the average particle density $\rho=n/V$: 

\begin{equation}
g(r) = \frac{\mathrm{d}N(V)}{\rho \mathrm{d}V} = \frac{1}{4 \pi r^2 \rho}\frac{\mathrm{d}N(r)}{\mathrm{d}r} \quad \text{with } V(r) = \frac{4}{3} \pi r^3
\label{eqn:RDF_def}
\end{equation}
 
Inserting \eqref{eqn:RDF_def} into \eqref{eqn:B22fromRDF} yields $B_{22}$ in terms of the cumulative number distribution function $N(r^\prime)$:  

\begin{equation}
B_{22}(r^{\prime}) = N_A \left(\frac{2\pi}{3}r^{\prime3}-\frac{N(r^{\prime})}{2\rho}\right)
\label{eqn:B22_from_N}
\end{equation} 
 
which obviates the need for numerical integration.

To be of practical use, the RDF method requires sufficient solute molecule numbers over the separation distance of interest, a prerequisite that is only fulfilled when the free energy landscape, {\it i.e.} the PMF, does not contain wells so deep that the Boltzmann-distributed solute molecules are effectively depleted from other regions. We found this limit to be on the order of -1~$RT$ (-2.5~kJ/mol), therefore if trial MD simulations for a given condition led to PMF minima not lower than -2.5~kJ/mol we employed the cumulative RDF method and otherwise the HEUS method, as detailed below.

Where appropriate, RDFs were obtained from MARTINI simulations of $n=420$ (glucose, sucrose) or $n=105$ (CDs, A2 glycan) randomly placed saccharide molecules in water-filled cubic boxes of volume [$V_b=(19\; \mathrm{nm})^3$]\changesfromto{, corresponding to concentrations of 100 and 25~mM respectively}.
These numbers, corresponding to concentrations of 100 and 25~mM respectively, ensure a sufficient degree of solute-solute interaction while being well below the solubility limit (except for $\beta$-CD which is soluble up to 16.5 mM in water).

The resulting trajectories of 1 to 10~$\mu$s were split into 200~ns segments and for every segment the cumulative saccharide-saccharide RDF (in terms of centers of mass) was computed using the GROMACS \it gmx rdf \rm  program\cite{Abraham201519}. The resulting curves were averaged to give the final RDF with standard deviation as an error estimate.

\paragraph*{Direct calculation of PMF}

In principle an MD simulation of two saccharide molecules could yield the PMF along the reaction coordinate of choice, that is the center of mass separation of the two molecules. However, in the case of large energy barriers along the reaction coordinate, (precluding efficient calculation of the RDF) mere Boltzmann sampling does not suffice to adequately sample the free energy landscape. In umbrella sampling\cite{Torrie1977} numerous individual MD simulations are run in parallel, each imposed with an additional biasing potential designed to restrain the reaction coordinate to a certain window with a given force constant. Afterwards the individual potentials are de-biased and used to reconstruct the final PMF. An additional improvement that mitigates entrapment in local minima is the stochastic exchange of the Hamiltonian, that is effectively the biasing potential, between neighboring windows (Hamiltonian exchange umbrella sampling, HEUS)\cite{heus}. 

In order to perform HEUS simulations, GROMACS v5.0.4 was patched with the PLUMED v2.2.1 plug-in\cite{Tribello2014604}.
The initial setup consisted of two solute molecules placed at maximum distance along the longest axis of an $8 \times 8 \times 12$ nm$^3$ simulation box. 
After  minimization and solvation, as described above, molecules were pulled towards each other while recording intermediate positions every 0.15~nm, until sterical clashes prevented further motion.
The centers of harmonic biasing potentials were set at these positions with uniform force constant of 500 $\mathrm{kJ/(mol \cdot nm^2)}$, spanning typically distances between 0.2 and 5~nm and yielding up to 32 independent umbrella windows.
Each window was further subjected to a short equilibration and subsequent production runs of 300 to 500~ns were carried out.
Every 1000 steps an attempt was made to exchange Hamiltonians of neighboring windows.
The center of mass positions and the corresponding value of biasing potential was recorded every 10 steps.

During pulling and production simulations, molecules were free to move along the long box axis and rotate, but the center of mass position along shorter axes was restrained with a force constant of \num{1E5} $\mathrm{kJ/(mol \cdot nm^2)}$ to prevent dumbbell-like rotation of paired molecules across the box.
Since this additional restraint was orthogonal to the direction in which the PMF was calculated and the rotational degrees of freedom were not affected, we expect this did not have an effect on the PMFs.
The same restraint was applied in all dimensions during equilibration steps.

In order to construct the PMF, HEUS simulations were de-biased using the Weighted Histogram Analysis Method (WHAM)\cite{wham} implemented in GROMACS\cite{gmxwham}.
Each simulation was repeated at least seven times with randomized molecule orientations in the starting configuration.
Average and standard deviation were estimated from a bootstrapping procedure (100 cycles) using replica histograms to reconstruct random PMFs.
The final PMFs were offset to zero energy at an intermolecular distance of 4~nm, where a plateau was reached for all molecules studied.

For the purpose of $B_{22}$ calculation, the integral in \eqref{eqn:B22_pmf} has to be finite:

\begin{equation}
B_{22}(r^{\prime})=-2\pi N_A \int_{0}^{r^{\prime}}[\exp(-\frac{W_2(r)}{RT})-1]r^2dr\label{eqn:B22fromPMF}
\end{equation}
 
and an $r^{\prime}=r^{\prime}_{0}$ has to be chosen large enough so that solute-solute interactions effectively vanish and $B_{22}(r^{\prime}_{0}) \rightarrow B_{22}(\infty)$. 
We found this consistently to be the case for $r^{\prime}>4$ nm, therefore we settled at a (conservative) $r^{\prime}_{0}$ = 5~nm in the course of this study.  
Numerical integration of \eqref{eqn:B22fromPMF} was performed using the trapezoid rule and a set of homemade scripts. Asymmetric 95\% confidence intervals were calculated, exploiting the fact that $B_{22}$ values obtained from bootstrapped PMFs were distributed log-normally\cite{lognormal}.
PMFs derived with both RDF and HEUS methods under the same conditions were found remarkably similar (fig.~\ref{fgr:B22}).

\subsection{Experimental methods}
\subsubsection{Biophysical characterization of A2 in aqueous solution}
A2 glycan (1~mg) was obtained as a powder from Ludger Ltd, Oxfordshire, UK (cat. no. CN-A2-SPBULK, batch no. B623-01).
The manufacturer had determined its purity, by hydrophilic interaction liquid chromatography of 2-aminobenzamide labeled A2 glycan, to be 92.0\% with 3.7\% and 0.6\% contaminants resulting from loss of one or both sialic acid moieties, respectively.

\subsubsection{Solution dispersity and hydrodynamic radius}
To assess the size distribution of dissolved A2 glycan molecules, solutions of 4 to 10~g/L in 100~mM NaCl were prepared, filtered (Whatman Anotop, 0.02~$\mu$m pore size), and dynamic light scattering (DLS) intensities were recorded in a DynaPro NanoStar instrument equipped with a calibrated 1.25~$\mu$L quartz cuvette (Wyatt Technology Corporation, Santa Barbara, CA).
Measurements were carried out at a single scattering angle $\theta= 90^{\circ}$, temperature $T = 300$ K, and incident vacuum laser wavelength $\lambda_{0} = 658$ nm.
The instrument's digital correlator (512 channels) was employed to compute $g^{(2)}(\tau)$, the normalized second order autocorrelation function (ACF) of the scattered light intensity:

\begin{equation}
g^{(2)}(\tau) = \frac{\langle I(t)I(t+\tau) \rangle}{\langle I(t) \rangle ^{2}}
\label{eqn:g2exp}
\end{equation} 

The second order ACF $g^{(2)}(\tau)$ can be related to the normalized first order electric field ACF $g^{(1)}(\tau)$ through\cite{goldin2002}:

\begin{equation}
g^{(2)}(\tau) = [g^{(1)}(\tau)]^{2} + \xi(\tau) + 1
\label{eqn:g2_g1}
\end{equation} 

with the experimental noise $\xi(\tau)$.
The field ACF $g^{(1)}(\tau)$ of a non-uniform (polydisperse) solution can theoretically be described\cite{Ogendal2016} as an integral of exponential decays corresponding to a distribution of different-sized scatterers:

\begin{equation}
\label{eqn:g1theo}
\begin{split}
g^{(1)}(\tau) &= \int_{0}^{\infty} B(\Gamma) \mathrm{e}^{-\Gamma \tau} \mathrm{d}\Gamma,\\
\textrm{with }\Gamma &= D_{t}q^{2},\\
\textrm{and }q &= 4 \pi \frac{n}{\lambda_{0}} \sin \frac{\theta}{2}
\end{split}
\end{equation}

with the translational diffusion coefficient $D_t$, the magnitude of the scattering vector $q$, the solution refraction index $n$, the vacuum wavelength of the incident light $\lambda_{0}$, and the scattering angle $\theta$. 
  
The DYNALS regularization method \cite{goldin2002} implemented in the DYNAMICS software (Wyatt Technology Corporation) was used to find a discretized approximation for $B(\Gamma)$ in \eqref{eqn:g1theo} whose coefficients represent the intensity fractions of the scattered light attributed to the different scatterers.
$B(\Gamma)$, and its discrete approximation, can be expressed in terms of hydrodynamic radii $R_{H}$ through the Stokes-Einstein relation:

\begin{equation}
R_{H} = \frac{k_{B} T}{6 \pi \eta_{0} D_{t}} = \frac{k_{B} T q^{2}}{6 \pi \eta_{0} \Gamma} \Leftrightarrow \Gamma = \frac{k_{B} T q^{2}}{6 \pi \eta_{0} R_{H}}
\label{eqn:Stokes-Einstein}
\end{equation}

where $\eta_{0}$ denotes the solvent viscosity (\num{1.019E-03} Pa$\cdot$s).

The mass fraction attributable to every scatterer $j$ was estimated from its intensity contribution $B_j(R_H,\theta)$ (eqns. \eqref{eqn:g1theo}, \eqref{eqn:Stokes-Einstein}) according to:
\begin{equation}
C_j \propto \dfrac{B_j(R_H, \theta)}{M_j(R_H) P(R_H,\theta)} 
\end{equation}
where $C_j$ is the mass concentration of scatterer $j$,$B_j(R_H, \theta)$ is the intensity of the scattered light for molecule $j$ with radius $R_H$ at scattering angle $\theta$, $M_j(R_H)$ is the molar mass,  and $P(R_H,\theta)$ is the scattering function (form factor). The Rayleigh-Gans approximation for random coils was chosen to estimate the angular dependence of $P(R_H,\theta)$, and it was assumed that $M_j(R_H) \propto R_H^2$ (Wyatt Technology Corporation, Technical Note 2004).

\subsubsection{Measurement of the second virial coefficient of the osmotic pressure, \texorpdfstring{$B_{22}$}{B22}}

Static ({\it i.e.} time-averaged) light scattering (SLS) intensities recorded from a series of solutions with varying concentration allow for the calculation of $B_{22}$. SLS intensities from the filtered A2 glycan solutions described in the DLS experiment were acquired using the SLS detector of the DynaPro NanoStar instrument. Temperature, scattering angle and laser light vacuum wavelength were kept the same ($\theta = 90^{\circ}$, $T = 300$~K, $\lambda_{0} = 658$~nm).

The time-averaged scattered light intensity $\langle I_{s} \rangle$ at scattering angle $\theta$ relates to the solute mass concentration $C$ and solute molecular mass $M_{W}$ through \cite{vanHolde1985, Ogendal2016}:
 
\begin{equation}
\begin{split}
\langle R_{\theta} \rangle \equiv \frac{ \langle I_{s}(\theta) \rangle}{\langle I_{0}(\theta) \rangle}\frac{r^{2}}{V_{s}(\theta)} &= \frac{4 \pi^{2}}{N_{A}}\frac{n_{0}^{2}}{\lambda_{0}^{4}} \left(\frac{\mathrm{d} n}{\mathrm{d}C}\right)^{2} M_{W} C S(\theta)\\
&\equiv K M_{W} C S(\theta)
\end{split}
\label{eqn:SLS}
\end{equation}

with $\langle I_{0} \rangle$ as time-averaged scattered light intensity of the solvent alone, the distance $r$ between detector and scattering volume $V_{s}$, the solvent refractive index $n_{0}$, the vacuum wavelength of the incident light $\lambda_{0}$, the solution refractive index $n$, and the structure factor $S(\theta)$.
The size of the refractive index increment $\frac{\mathrm{d} n}{\mathrm{d}C}$ can be assumed constant over the concentration range under examination, and here we used the empirically found value of 0.145 ($\pm$ 0.005) mL/g for polysaccharides\cite{Ogendal2016}.
The structure factor $S(\theta)$ describes the phase relation of the light scattered from the molecules in $V_{s}$. Ideal solute molecules (no intermolecular forces) scatter with completely uncorrelated phases, and $S(\theta)$ becomes unity. 
Non-ideal solute molecules influence each other (excluded volume, attractive or repulsive intermolecular forces) thus having a structure factor $S(\theta)$ deviating from one. 
For non-ideal molecules that are much smaller than the wavelength of the incident light, as is the case for A2 glycan in the setup described, it has been shown \cite{Tanford1961,Ogendal2016} that

\begin{equation}
\frac{1}{S} = 1+2A_{2}M_{W}C+3A_{3}M_{W}C^{2}+\dots
\label{eqn:S}
\end{equation}

with $A_{2}$, $A_{3}$, \dots as second, third, and higher coefficients of the virial expansion of the osmotic pressure $\Pi$ expressed in terms of the solute mass concentration $C$:

\begin{equation}
\frac{\Pi}{RT} = \frac{1}{M_{W}}C + A_{2}C^{2} + A_{3}C^{3} + \dots 
\label{eqn:osmoticP_mass}
\end{equation} 

Inserting eq. \ref{eqn:S} in eq. \ref{eqn:SLS} gives after rearrangement

\begin{equation}
\frac{KC}{R_{\theta}} = \frac{1}{M_{W}} + 2A_{2}C + 3A_{3}C^{2} + \dots 
\label{eqn:Zimm_light}
\end{equation}

The third and higher virial coefficients can be neglected in the limit of low concentrations, resulting in a linear relationship of $C \to \frac{KC}{R_{\theta}}$ in \eqref{eqn:Zimm_light} which yields the molar mass $M_{W}$ from the y-intercept and $A_{2}$ from the slope.   
The conversion of the mass concentration based virial coefficients $A_{i}$ into the mole concentration based coefficients $B_{2i}$ is achieved by comparing eqs. \ref{eqn:osmoticP_mole} and \ref{eqn:osmoticP_mass}:

\begin{equation}
B_{22} = A_{2}M_{W}^{2}, \quad B_{23} = A_{3}M_{W}^{3}, \quad \dots 
\label{eqn:A2_B22}
\end{equation}

\section{RESULTS}

\subsection{Selection of saccharides\changesfromto{none}{, simulation conditions and water model}}
We based our analysis on a complex biantennary N-glycan, hereafter referred to as A2 glycan (fig. \ref{fgr:mapping}A) which represents a structure commonly found as asparagine-linked N-glycan in vertebrate glycoproteins\cite{Royle2008,Hua2013,gao_microfluidic_2015} and can be thought of as a prototype of protein glycosylation.
It consists of a conserved pentasaccharide core that is substituted by two identical trisaccharide antennae, each terminating with sialic acid residues.
Sialic acid carries a carboxyl group which is deprotonated and thus negatively charged at physiological pH.
A2 glycan is commercially available as a product purified from natural resources.
In addition, to test the hypothesis that the aggregation behavior of polysaccharides in MARTINI depends on their size, we \changesfromto{added to our analysis a range of}{included in our analysis four extensively studied} saccharides for which experimental values of $B_{22}$ are known: The monosaccharide glucose\footnote{Unless otherwise stated monosaccharides are referred to as in their D-pyranose form.}\changesfromto{ and}{,} the disaccharide sucrose ($\alpha$-glucosyl-(1$\rightarrow$2)-$\beta$-fructofuranose), \changesfromto{which in animals and plants are important energy sources and have been extensively studied.}{and the cyclic oligosaccharides $\alpha$- and $\beta$-CD which are six- and seven-membered cyclic oligomers of $\alpha$1$\rightarrow$4-linked glucose.}
\changesfromto{Within the MARTINI framework glucose is structurally identical to the other biologically relevant monosaccharides galactose and mannose. Finally, we study two long-known cyclic oligosaccharides of $\alpha$1$\rightarrow$4-linked glucose, namely the six-membered $\alpha$-CD and the seven-membered $\beta$-CD.}{}
\changesfromto{none}{
Since the aggregation propensity of saccharides in aqueous condition depends on the balance of solute-solute, solute-water, and water-water interactions, the choice of the water model is of particular importance. 
Currently, three water models exist in MARTINI: 
First, the standard water model\cite{Marrink2007}, in which groups of four H$_2$O molecules are represented by a single, uncharged P$_4$ bead interacting solely through LJ potentials ($\sigma=0.47$ nm, $\epsilon=5.0$ kJ/mol towards other water particles). 
Second, antifreeze water, a mixture of standard water and antifreeze particles (usually 10\%)\cite{Marrink2007} to prevent freezing at 280-300~K. The antifreeze particles behave as water particles except for interaction with the standard water particles ($\sigma=0.57$ nm, $\epsilon=5.6$ kJ/mol).
Third, a polarizable water model\cite{Yesylevskyy2010} in which the central, uncharged bead is connected to two partially charged ($\pm{0.46\;e}$) non-LJ interacting beads and the water-water LJ interaction has been reduced compared to the standard water model ($\sigma=0.47$ nm, $\epsilon=4.0$ kJ/mol).\\
In our simulations we chose a temperature of 300~K to ensure comparability of calculated $B_{22}$ values with experimental results which have typically been acquired at 298~K. At this temperature, we frequently observed freezing of MARTINI simulations of very diluted systems containing standard water as a solvent. This precluded the HEUS approach to calculate the PMF (see Methods) and thus made calculations of $B_{22}$ for strongly interacting particles impossible with the standard water model. Nevertheless the standard water model in MARTINI is far more often used than the polarizable water model, which is why we chose to use antifreeze water in all our MARTINI simulations. The same water model has been employed by \citeauthor{Stark2013}\cite{Stark2013} which makes their findings on the aggregation propensity of proteins directly comparable to ours.} 

\subsection{Spurious aggregation of A2 glycan}
\begin{figure}[ht!]
\includegraphics[width=0.45\textwidth]{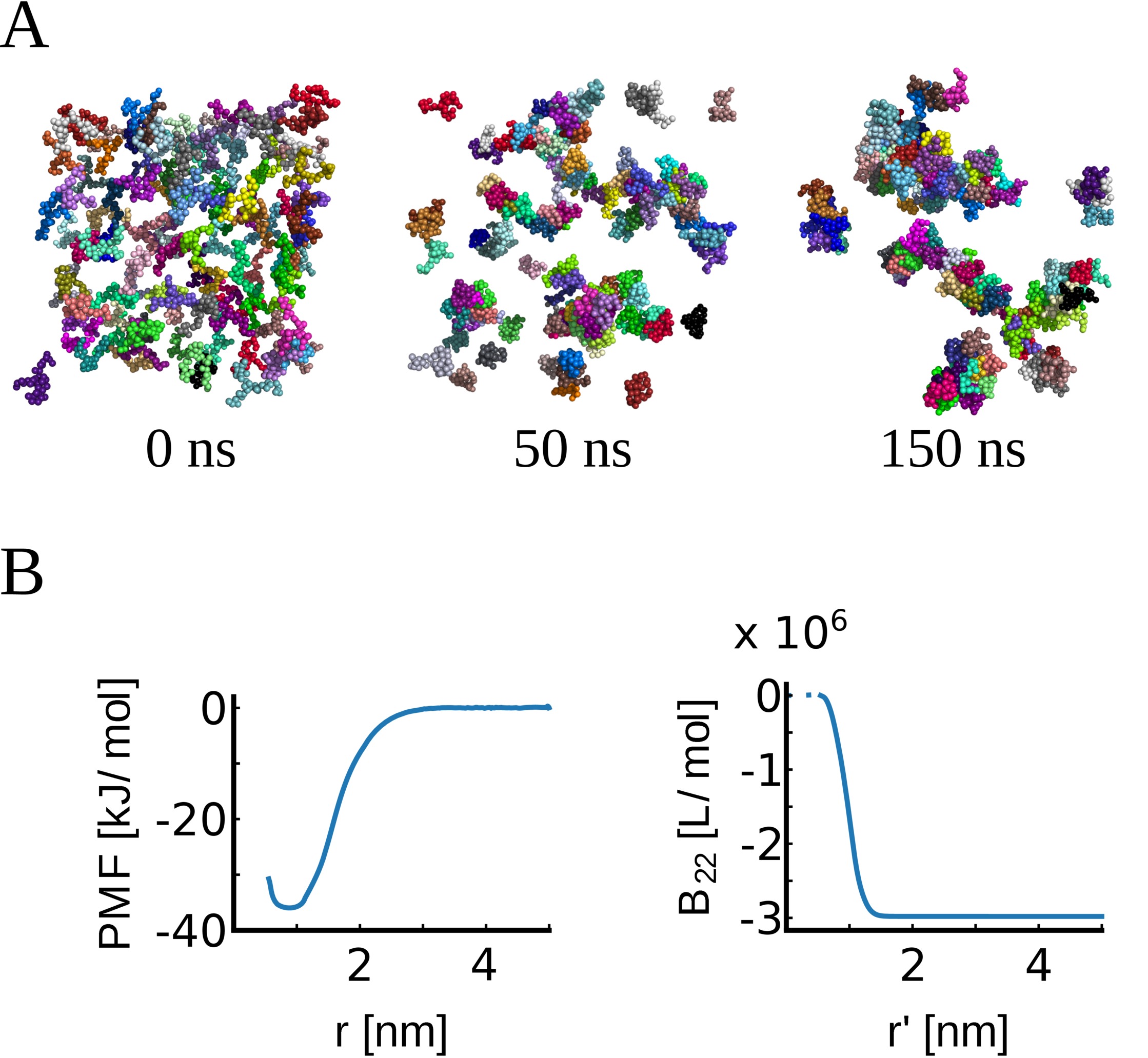}
 \caption{A: Snapshots from a MARTINI simulation of 105 A2 glycan molecules. Each molecule is represented by a separate color. Times reported represent actual simulation time, without correction due to changed diffusional dynamics\cite{monticelli_martini_2008}.
 B: PMF for A2 glycan in the same conditions and corresponding calculation of the second virial coefficient of the osmotic pressure, $B_{22}$. Note the depth of the potential well and the extremely low value of $B_{22}$.}
 \label{fgr:aggregation}
\end{figure}

We hypothesized that a potential discrepancy between MARTINI and experimental aggregation would increase with saccharide size; consequently, we chose the largest glycan in our selection, A2 glycan, for an initial test for intermolecular interaction in MARTINI.
To this end, 105 A2 glycan molecules were placed in a cubic (19~nm)$^3$ box filled with MARTINI water \changesfromto{with}{and} 100~mM NaCl and 10~mM $\mathrm{CaCl_2}$. The A2 glycan is expected to be readily soluble at this concentration (25~mM or 55.6~g/L), as {\it e.g.} dextrans (branched polymers of mainly $\alpha$1$\rightarrow$6-linked glucose) of medium to high molecular weight easily dissolve in water up to 400~g/L\cite{Ioan2000}. 
In the simulation however, we observed a striking aggregation behavior, resulting in all A2 glycan molecules clumping together within a few tens of nanoseconds (fig.~\ref{fgr:aggregation}A). Furthermore, evolution of the system for a total of 1~$\mu$s revealed not a single dissociation event, which would limit A2 glycan solubility to less than one molecule per volume of the simulation box, {\it i.e.} $<$~0.25~mM. As we demonstrate in the experimental results section below, A2 glycan is readily soluble at concentrations up to 4.5~mM, and we therefore suspected that, in line with our hypothesis, \changesfromto{attractive sugar-sugar interactions were indeed severely overestimated}{the balance of non-bonded forces was severely biased towards promoting saccharide attraction}.

\changesfromto{none}
{Interestingly, we observed very similar aggregation of A2 glycan molecules with the other two water models, {\it{i.e.}} standard water and polarizable water\cite{Yesylevskyy2010} (fig.~\ref{SI:ions_WF}C,D), suggesting that we found a general problem in MARTINI rather than an isolated issue with a particular water model.}

\begin{table*}
  \caption{Experimental values of $B_{22}$ for diverse saccharides}
  \label{tbl:B22exp}
  \begin{tabular}{p{2.7cm}llllll}
    
 Molecule             & Mw [g/mol]        &  Solvent       &   T [K]    &   $B_{22}$ [L/mol]& $A_2$ [mol L/g2] &  Ref.\\
 \hline
glucose               & 180.16            & water          &  298.15    &  0.117            & \num{3.61e-6}   & \citenum{Stigter1960}    \\
cellobiose            & 342.30            & water          &  298.15    &  0.267            & \num{2.28e-6}   & \citenum{Herrington1983} \\
sucrose               & 342.30            & water          &  298.15    &  0.305            & \num{2.60e-6}   & \citenum{Stigter1960}    \\
trehalose             & 342.30            & water          &  295       &  0.51             & 
\num{4.3e-6}    & \citenum{Davis2000}      \\
$\alpha$-CD     & 972.85            & water          &  298.15    &  0.830            & \num{6.57e-7}   & \citenum{Terdale2006}    \\
$\beta$-CD      & 1135              & water          &  298.15    &  6.296            & \num{4.89e-6}   & \citenum{Dagade2004}     \\
A2 glycan             & 2224              & 0.1M NaCl      &  300       &  46               & \num{9.3e-6}   & this work             \\
dextran               & 9000              & 0.01M NaN$_3$  &  293.15    &  {60.7}    & {\num{7.49e-7}}   & \citenum{Ioan2000}       \\   
dextran               & 37400             & 0.01M NaN$_3$  &  293.15    &  {590}    & {\num{4.22e-7}}   & \citenum{Ioan2000}       \\   
dextran               & 59000             & 0.01M NaN$_3$  &  293.15    &  {1590}    & {\num{4.56e-7}}   & \citenum{Ioan2000}       \\   
dextranT2000          & \num{1.58e6}      & 0.03M NaCl     &  302.05    &  \num{7.5e+5}     & \num{3.0e-7}    & \citenum{Schaink2007}    \\
   \end{tabular}
\end{table*}

To quantify our findings, we calculated the potential of mean force (PMF, see Methods) of A2 glycan using HEUS (Hamiltonian Exchange Umbrella Sampling) and computed $B_{22}$ (fig.~\ref{fgr:aggregation}B). 

Corroborating our qualitative findings \changesfromto{(fig.~\ref{fgr:aggregation}A)}{}we found a pronounced well of -36~kJ/mol at a distance of 0.7-1~nm and concomitantly $B_{22}$ converged to \num{-1.3E06}~L/mol. Experimental $B_{22}$ values for other saccharides have all been found positive (cf. table \ref{tbl:B22exp}), so we sought to measure $B_{22}$ of A2 glycan experimentally to scrutinize the prediction from MARTINI.

\subsection{Experimental determination of \texorpdfstring{$B_{22}$}{B22} of A2 glycan}
$B_{22}$ values have been determined for the smaller saccharides (glucose, cellobiose, sucrose, $\alpha$-, and $\beta$-CD)\cite{Stigter1960, Herrington1983, Dagade2004, Terdale2006} and some high molecular weight dextrans \cite{Ioan2000, Schaink2007}, but to our knowledge not for A2 glycan or any similar structure. To elucidate the thermodynamic properties of A2 glycan in aqueous solution with only low amounts of available A2 glycan (1~mg), we performed dynamic and static light scattering experiments which can be conducted with sample volumes as small as 1.5~$\mu$L.

\begin{figure*}
 \includegraphics[width=0.9\textwidth]{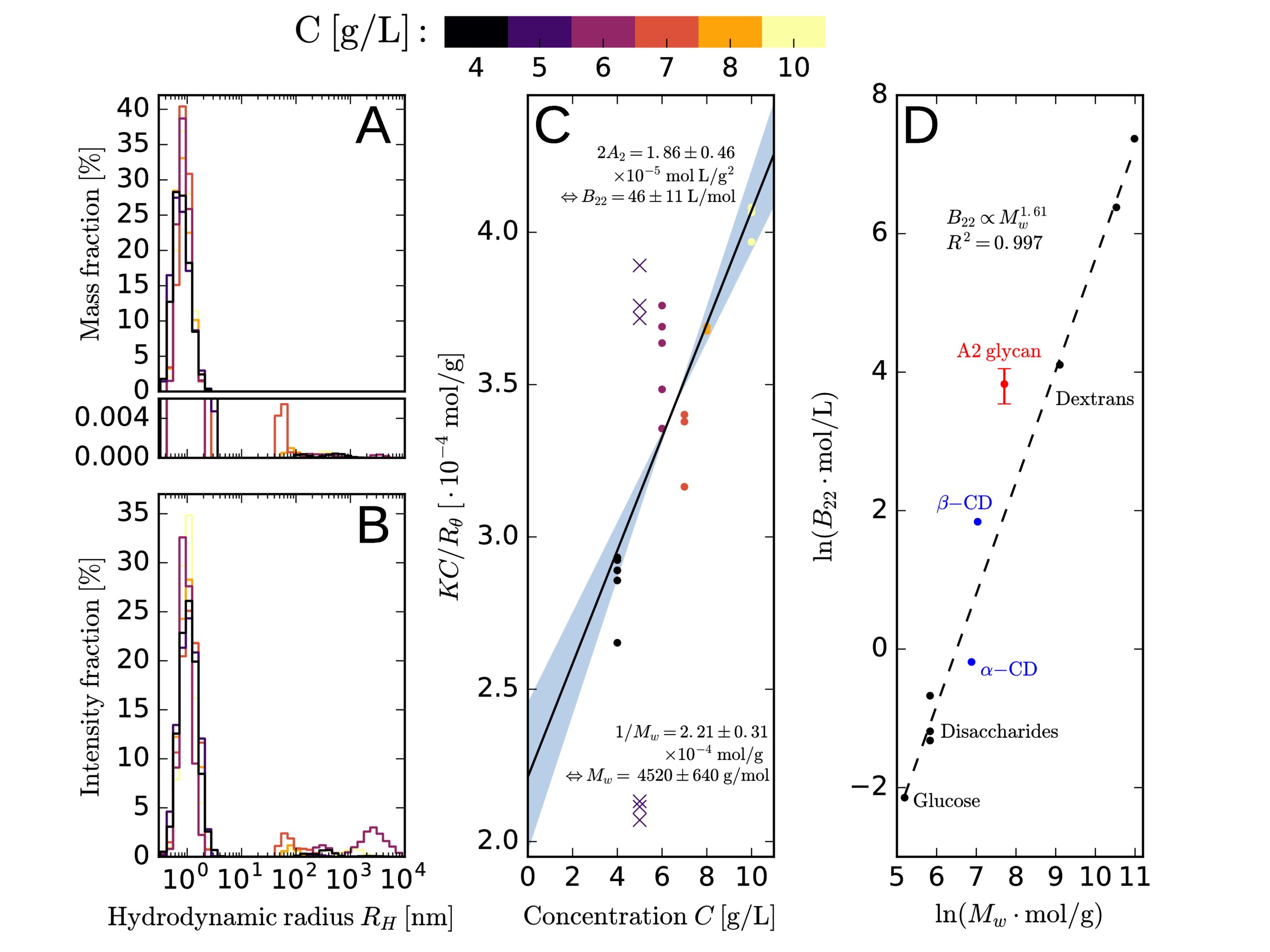}
 \caption{A, B: Average size distribution of scattering particles in filtered solutions of A2 glycan in 0.1 M NaCl obtained from dynamic light scattering experiments ($n$ = 27, $C_{\mathrm{A2 glycan}}$ = 4-10 g/L, see color code). The peak centered around 1.2-1.3 nm accounts for more than 99.9\% of the estimated mass fraction (A) and more than 96\% of the intensity fraction (B). C: Molecular mass, $M_{W}$, and second virial coefficient of the osmotic pressure, $B_{22}$, obtained from a linear fit to a Zimm plot of time-averaged (static) scattering light intensities measured from the same solutions as in panels A, B. Values at 5 g/L (x labels) were considered outliers due to their obvious non-Gaussian distribution and excluded from the fit. Numeric errors and blue shading denote 95\% confidence interval of the linear fit parameters. \changesfromto{none}{D, Empirical $B_{22}(M_w)$ relationship for linear uncharged saccharides (black dots and black dashed line). Cyclodextrins (CDs, blue) and A2 glycan (red) deviate from this trend. Error bars denote 95\% confidence interval.}}
 \label{fgr:B22exp}
\end{figure*}

A2 glycan was dissolved in 100~mM NaCl to mimic physiological conditions. As a prerequisite to $B_{22}$ measurements, we determined the dispersity of the A2 glycan solution by dynamic light scattering. The hydrodynamic radius of the main species of A2 glycan solutions (4-10~g/L) was \changesfromto{measured as}{found to be} 1.2-1.3~nm, which is in very good agreement with an A2 glycan monomer (fig.~\ref{fgr:B22exp}A, B). Only tiny amounts of larger aggregates were observed (fig.~\ref{fgr:B22exp}A), suggesting that A2 glycan dissolved in water into an essentially uniform (monodisperse) solution of monomers. Notably, A2 glycan dissolved entirely up to a concentration of 10~g/L (4.5~mM), well above the upper solubility limit of 0.25~mM suggested by the MARTINI simulation.

Static light scattering analysis of A2 glycan solutions of different concentrations ($C=4..10$~g/L in 100~mM NaCl) yielded a linear relationship (fig.~\ref{fgr:B22exp}C) between $KC/R_{\theta}$ (see Methods for details) and $C$. From the y-intercept of the linear fit in fig.~\ref{fgr:B22exp}C a molecular mass $M_{W}$ of $4500\pm{700}$~g/mol \changesfromto{none}{(95\% confidence interval (CI))} was obtained which is around twice the theoretical $M_W$ of a monomer (2224~g/mol). From a chemical perspective it seems unlikely that A2 glycan spontaneously dimerizes in aqueous solution, which is why we attribute this discrepancy to systematic errors in $C$. The slope of the fit directly relates to $B_{22}$ of A2 glycan and was found to be ${46}\pm{11}$~L/mol \changesfromto{none}{(95\% CI)}. It is noteworthy that this clearly positive value indicates a net repulsive interaction between A2 glycan molecules in solution which is strongly opposing the behavior of A2 glycan in the MARTINI simulation. Table \ref{tbl:B22exp} \changesfromto{}{and fig.~\ref{fgr:B22exp}D} show that the measured $B_{22}$ value for A2 glycan fits well into the series of known $B_{22}$ values for other saccharides. \changesfromto{which have all been found positive}{The $B_{22}$ of A2 glycan is found to be more positive than would be expected for an uncharged saccharide of the same mass, which can be explained by an additional repulsive interaction due to the two negative charges of A2 glycan}. 
Notably, the physiological salt concentration employed in our MD simulations had negligible effect on $B_{22}$ (fig. \ref{SI:ions_WF}).

Taken together, we found our hypothesis confirmed that \changesfromto{attractive intermolecular forces between A2 glycan molecules are overestimated}{at least for A2 glycan the non-bonded force balance was biased towards promoting aggregation} in MARTINI. We next investigated ways to adjust the MARTINI force field parameters to make it better reproduce experimental $B_{22}$ values.

\subsection{Ways to improve the MARTINI force field}
Microscopically, $B_{22}$ reflects the relative strengths of solute-solute, solute-solvent, and solvent-solvent non-bonded interactions, which are \changesfromto{composed of}{modeled by} a Coulomb and an LJ interaction potential \changesfromto{none}{in MARTINI}. The overestimation of the aggregation propensity occurs irrespective of electrical charge (all beads are uncharged except for two negatively charged beads in A2 glycan), necessitating adjustments to the LJ interaction.
\changesfromto{The water-water interaction in MARTINI is a careful compromise between the reproduction of the experimental Gibbs energy of vaporization and the stability of the liquid phase at 300~K\cite{Marrink2007}. Clearly, this interaction is central to the MARTINI force field and, despite known insufficiencies\cite{Marrink2007}, we found its re-parametrization was beyond the scope of this study.   
MARTINI has been parametrized to reproduce partition coefficients of small molecules in various water/organic solvent systems. Theoretical partition coefficients have been obtained from Gibbs energies of solvation in MARTINI with water, chloroform, ether, hexadecane, or octanol as solvent. These partition coefficients would have changed upon selectively modifying the saccharide-water interaction strength, thus we kept these interactions unaltered.  
These methodological constraints narrow the selection of tunable parameters to}{Specifically, we sought a modification that would retain central force field properties, {\it i.e.} the mapping rules and the partitioning behavior between water and apolar solvents, to circumvent a complete reparametrization of MARTINI.
Therefore the water-water and water-solute interactions were kept unchanged, leaving} $\sigma_{ij}$ of the saccharide beads and $\epsilon_{ij}$ of the saccharide-saccharide interactions, both of which are expected to influence $B_{22}$, as tunable parameters.
However, tests of scaling $\sigma_{ij}$ for all interactions involving saccharide beads either had insufficient effect on $B_{22}$ or, due to the potential shift of the LJ potentials, impacted $\epsilon_{ij}$ as well (fig.~\ref{SI:B22_sigma}).
We therefore left $\sigma_{ij}$ unmodified and instead analyzed the effect of a reduction of the interaction strength, $\epsilon_{ij}$, of saccharide-saccharide interactions on $B_{22}$.
\changesfromto{none}{
It has to be emphasized that this approach is only possible because solute-solute interactions can be expected to have a minor effect on partitioning coefficients, which depend mostly on solute-solvent and solvent-solvent interactions and were a key feature in the parametrization of the MARTINI force field\cite{Marrink2007}.
It does not imply that these interactions are the only cause for an imbalance in the non-bonded interactions. 
In fact, the water model in MARTINI is somewhat problematic and a known imbalance exists between solute-water and water-water interactions (see discussion below). 
Other approaches like improving the water model or changing water-solute interactions are possible routes to establish a better force balance. However, this would essentially mean a complete reparametrization of the MARTINI force field which is beyond the scope of this study.}
   
\subsection{Scaling of solute-solute interactions in MARTINI: A2 glycan and glucose}

In order to systematically investigate the effect of reducing $\epsilon$, we defined a scaling parameter~$\gamma$:

\begin{equation}
\epsilon_{ij,scaled}= 2 \text{ kJ/mol} + \gamma(\epsilon_{ij,original} - 2 \text{ kJ/mol})
\label{eqn:gamma_def}
\end{equation}

\begin{figure*}[ht!]
  \includegraphics[width=0.9\textwidth]{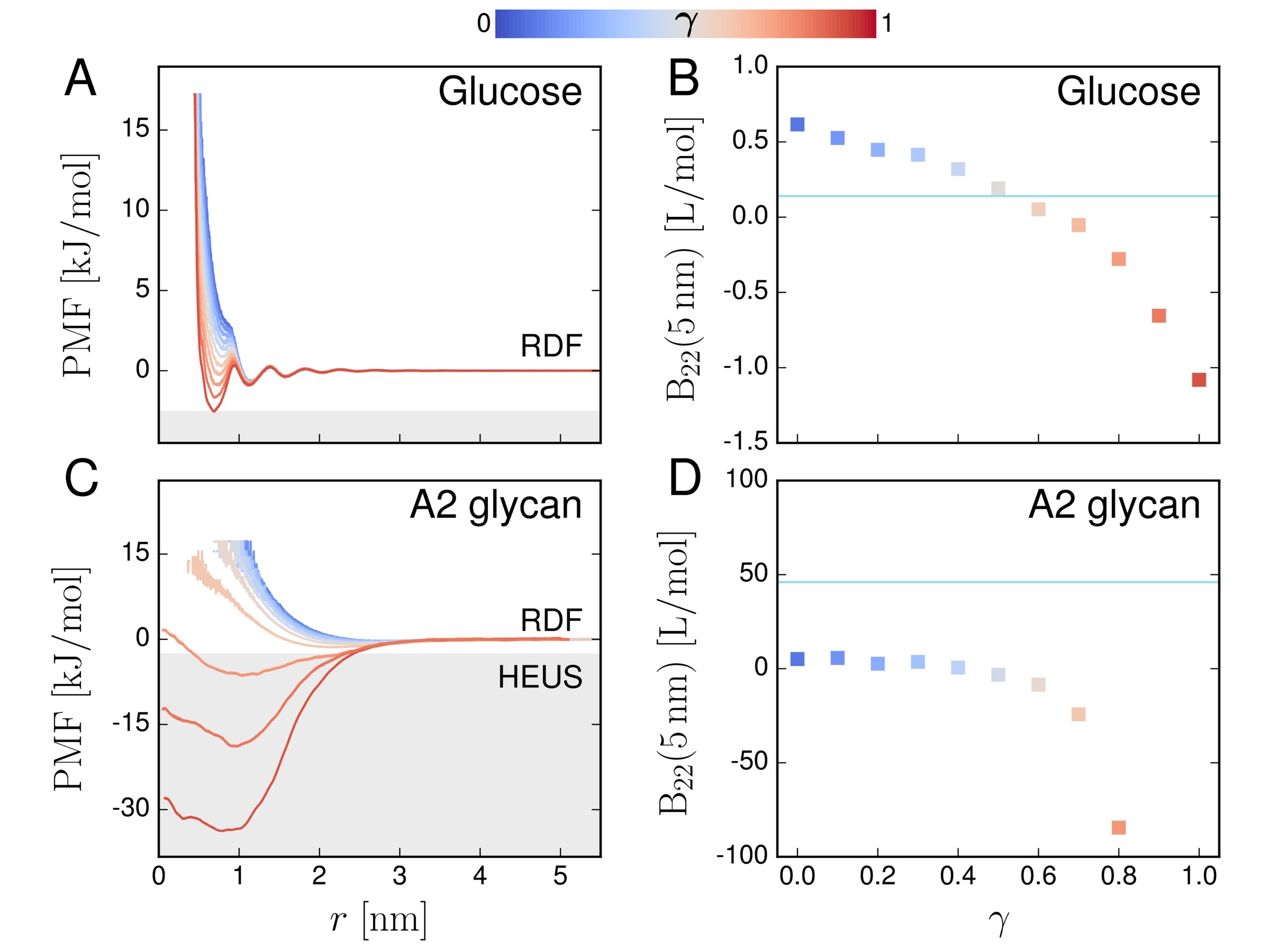}
  \caption{
Dependence of PMF (A, C) and $B_{22}$ (B, D) on the LJ scaling factor $\gamma$ for glucose (A, C) and A2 glycan (B, D). For A2 glycan with $\gamma>0.7$ (C, grey-shaded area) PMFs were constructed from HEUS simulations, in all other cases PMFs were generated from RDFs using eq.~\eqref{eqn:Boltzmann}. The color bar indicates the value of $\gamma$. In panels (B, D) $B_{22}(r^\prime)$ was evaluated at $r^\prime=5 \;\mathrm{nm}$ according to eqs.~\eqref{eqn:B22fromPMF} and \eqref{eqn:B22_from_N}. The blue horizontal line in indicates experimental $B_{22}$ values for glucose and A2 glycan (0.117 and 46 L/mol, respectively). See methods for details.}
  \label{fgr:gamma}  
\end{figure*}

Thus, $\epsilon_{ij,scaled}$ changes linearly in the interval [0,1] from 2 kJ/mol, the weakest LJ interaction level in MARTINI, to its original value $\epsilon_{ij,original}$. This approach follows the work of \citeauthor{Stark2013} who defined an analogous scaling factor $\alpha$ to modify protein-protein interaction levels\cite{Stark2013}. 
Figure \ref{fgr:gamma} shows the effect of varying $\gamma$ between 0 and 1 on the PMF and $B_{22}$ for glucose and A2 glycan. In the case of glucose, unscaled MARTINI ($\gamma=1$) yields a PMF with a potential well of -2.5~kJ/mol with a concomitant $B_{22}$ value of -1.1~L/mol. The implied \changesfromto{none}{net} attractive intermolecular interaction disagrees with the experimental prediction of a weak repulsion ($B_{22} = 0.117\;\mathrm{L/mol}$\cite{Stigter1960}). The depth of the potential well in the PMF flattens with decreasing $\gamma$ (fig.~\ref{fgr:gamma}A) and $B_{22}$ reaches good agreement with experiment at $\gamma=0.5$ (fig.~\ref{fgr:gamma}B).
Similarly, the deep potential well of the PMF of A2 glycan in unscaled MARTINI (-35~kJ/mol, fig.~\ref{fgr:gamma}C) flattens quickly with decreasing $\gamma$, shifting $B_{22}$ by six orders of magnitude from \num{-1.3E06}~L/mol to -3~L/mol at $\gamma=0.5$. However, the experimental $B_{22}$ value of 46~L/mol is never reached; even with lowest LJ interaction potentials ($\gamma=0$, $\epsilon_{ij}=2\;\mathrm{kJ/mol}$) $B_{22}$ reaches not more than $\approx$ 7~L/mol (fig. \ref{fgr:gamma}D). The value of $\gamma=0.5$ thus constitutes a compromise between reproduction of physical $B_{22}$ values and minimization of the changes to the original MARTINI force field parameters.

\changesfromto{none}{Again, we wish to emphasize that this is strictly valid for the antifreeze water model only. Qualitatively, the other two water models (standard water without antifreeze particles and polarizable water) seem to overestimate the aggregation propensity of A2 glycan, too (fig.~\ref{SI:ions_WF}C, D), however the optimal value of the scaling factor $\gamma$ will most likely be different. Quantitative tests with glucose and A2 glycan suggest for example that a scaling factor of $\gamma=0.8$ might suffice for standard water without antifreeze particles (fig.~\ref{SI:ions_WF}A, B).}

\subsection{Extension to sucrose and cyclodextrins}
\begin{figure*}
\includegraphics[width=0.9\textwidth]{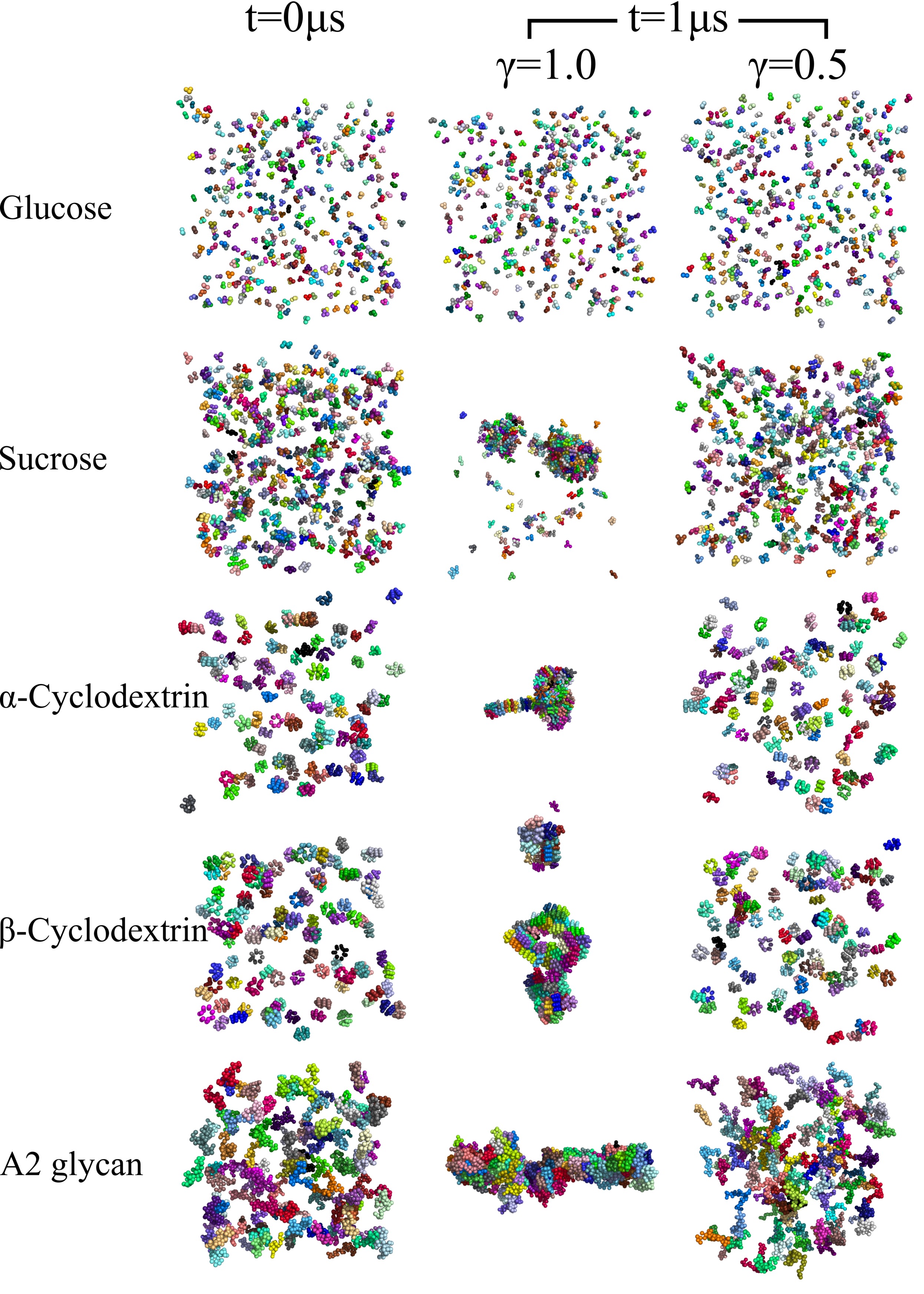}
\caption{Snapshots of unscaled ($\gamma=1.0$) and scaled ($\gamma=0.5$) MARTINI simulations for studied saccharide solutions in water. Each molecule is represented in different color. }
\label{fgr:snap_all}
\end{figure*}

\begin{figure*}[t]
  \includegraphics[width=0.99\textwidth]{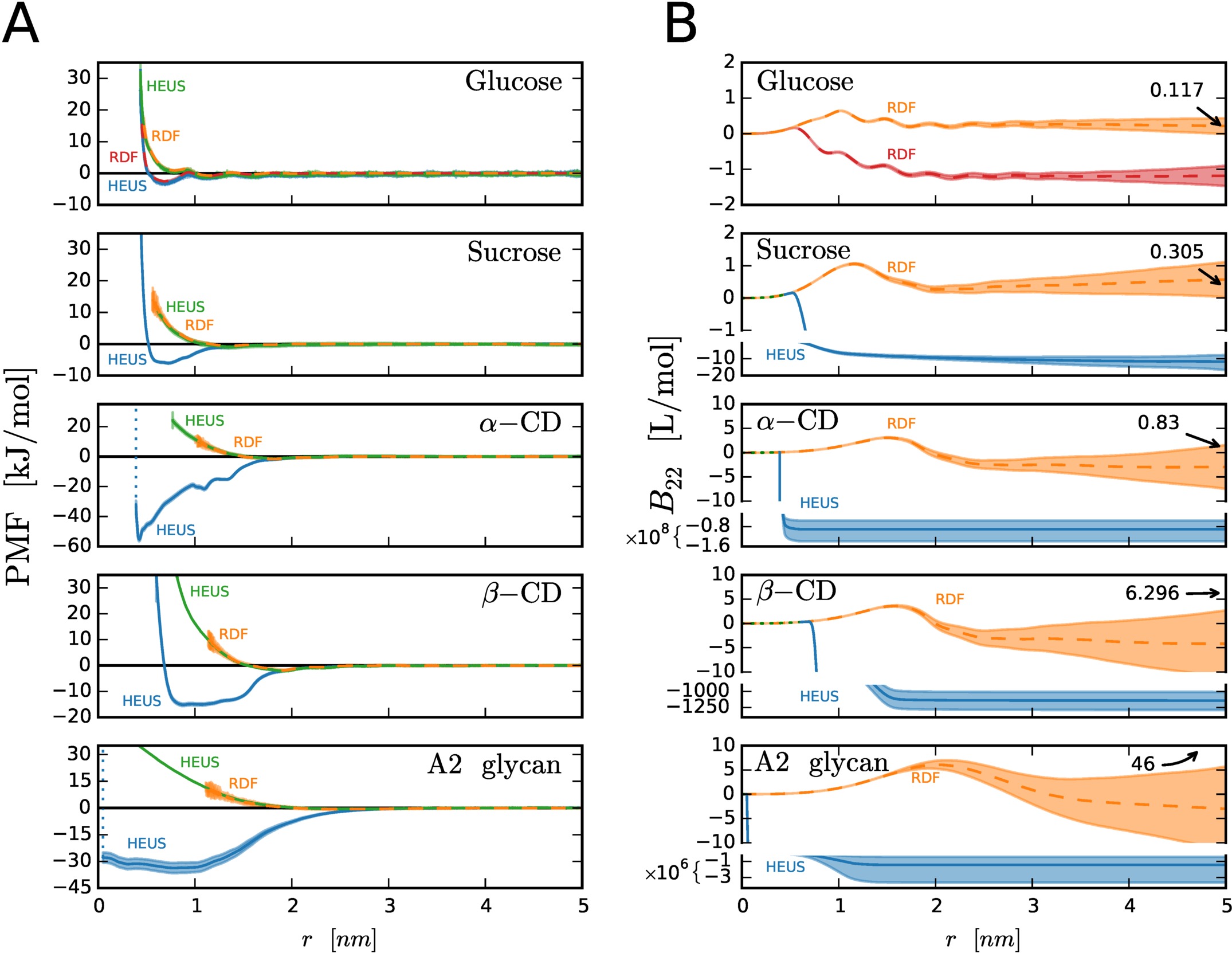}
  \caption{PMF and $B_{22}$ of saccharides with original (blue and red lines) and scaled ($\gamma=0.5$; green and orange lines) LJ interactions. Additional labels (HEUS, blue and green; RDF, red and orange) indicate the method of PMF calculation.
A: PMF plotted as a function of inter-particle distance. B: $B_{22}$ as a function of running upper limit of integration $r^\prime$. The 95\% confidence interval is shown as a semi-transparent envelope. Split y-axes were necessary to accommodate both curves in a single graph. Arrows indicate experimental values of $B_{22}$ in L/mol, cf. Table~\ref{tbl:B22exp}.}
  \label{fgr:B22}
\end{figure*}

To determine what effect a reduced LJ interaction strength would have for intermediate sized saccharides, we compared unmodified MARTINI simulations for sucrose, $\alpha$-, and $\beta$-CD (at 25~mM) with their scaled counterparts ($\gamma$=0.5). In unscaled MARTINI nearly all CD molecules and most of the sucrose molecules formed clusters within the first hundreds of nanoseconds of simulation time (fig.~\ref{fgr:snap_all}). The discrepancy with the aqueous solubility for these saccharides (sucrose: 1.97~M, $\alpha$-CD: 121~mM\cite{Pharr1989}, $\beta$-CD: 16.3~mM\cite{Pharr1989}; 25~$^{\circ}$C) indicates already \changesfromto{that saccharide-saccharide interaction strengths were overestimated in all these cases}{a clear overestimation of the aggregation propensity}. The PMF well depth in unscaled MARTINI varied from -6 to -55~kJ/mol (fig.~\ref{fgr:B22}A), whereas the energy of thermal fluctuations (RT) at 300~K equals to 2.5~kJ/mol. Clearly, the most probable state for an assembly of such molecules is strongly bound. Consequently, the corresponding $B_{22}$ values, even though endowed with significant uncertainties, all point to a very strong aggregation propensity (fig.~\ref{fgr:B22}B). Conversely, scaling down the LJ interaction strength with $\gamma$=0.5 resulted in fully disperse solutions for all saccharides over a time course of up to 10~$\mu$s (fig.~\ref{fgr:snap_all}). 
The derived and experimental $B_{22}$ values have been gathered in table~\ref{tbl:B22}.

\begin{table}
  \caption{Comparison of simulated and experimental $B_{22}$ values for the five studied saccharides. RDF-derived values (symmetric errors) represent means, PMF-derived values (asymmetric errors) medians. Errors always denote 95\% confidence interval ($\pm{2}$ SD).}
  \label{tbl:B22}
  \begin{tabular}{p{2.7cm}llll}

                         & \multicolumn{3}{c}{$B_{22}$ (L/mol)}  \\   
                         & \multicolumn{2}{c}{simulated}                                            & experimental\\
Saccharide               & $\gamma=1.0$                            & $\gamma=0.5$\\
 \hline
glucose                  &  ${-1.2}\pm{0.3}$                       &  \phantom{$-$}${0.22}\pm{0.23}$&  0.117\\
sucrose                  &  \tol{-11}{4}{6}                        &  \phantom{$-$}${0.57}\pm{0.55}$&  0.305\\
$\alpha$-CD    &  \tol{-8.8}{3.3}{5.1}$\times$\num{e07}  &  $-3.0\pm{4.6}$                                      &  0.830\\
$\beta$-CD     &  \tol{-1.1}{0.2}{0.2}$\times$\num{e03}  &  $-4.2\pm{6.9}$                &  6.296\\
A2 glycan                &  \tol{-1.3}{0.8}{2.4}$\times$\num{e06}  &  $-3.0\pm{8.7}$                &  $46\pm{11}$\\
   \end{tabular}
\end{table}

Clearly, unmodified MARTINI \changesfromto{fails to capture correct solute-solute interactions}{overestimates the aggregation propensity} in all studied cases.
The discrepancies generally escalate with growing $M_w$, corroborating our hypothesis about their size dependence.
Interestingly, whereas experimental values of $B_{22}$ of saccharides indicate a direct correlation between size and increased intermolecular repulsion, MARTINI simulations point to a strong inverse relationship, fostering the idea that a small overestimation in individual $\epsilon_{ij}$ accumulates in large molecules, leading to the observed net attractive forces.
Scaling of LJ interaction strengths led to a very good agreement of $B_{22}$ with corresponding experimental values for glucose and sucrose. 
For $\alpha$- and $\beta$-CD, agreement is still acceptable although $B_{22}$ values remained slightly negative, pointing to a weak tendency to aggregate. 
For A2 glycan, despite missing the experimental $B_{22}$ value, an improvement over orders of magnitude has been achieved. Notably, the obtained $B_{22}$ values for the CDs and A2 glycan seem to be identical, pointing to a possible size-independent upper limit of the proposed correction (see below for discussion).
 
\changesfromto{Intriguingly, one of the aggregation-enhancing factors seems to be the antifreeze particles (10\% of the water particles) that are necessary in MARTINI to keep water in its liquid state\cite{Marrink2007}. Water in MARTINI has originally been parametrized without antifreeze particles, and their addition reduces the average strength of water-water interactions. When antifreeze particles were removed, the $B_{22}$ value of A2 glycan raised from -3.0 to +9.2 L/mol in scaled MARTINI, illustrating the dependence of $B_{22}$ on the relative strength of solvent-solvent and solute-solvent interactions. Since in the current state antifreeze particles are unavoidable (especially in systems with low solute concentration, like the HEUS simulations), we present our findings in the default condition, {\it i.e.} with antifreeze water.}{}  

The limitations notwithstanding, even if a precise match between predicted and experimental $B_{22}$ could not be met, the aggregation behavior between $\gamma=1.0$ and $\gamma=0.5$ is critically different. Shallow PMF well depths account for an observed lack of aggregate formation, allowing for more realistic simulation of saccharides and/or glycosylated macromolecules. \changesfromto{An update of MARTINI water and further refinement of the non-bonded parameters will be, however, necessary to fine-tune solute-solute interactions}{}.

\subsection{Validation of the proposed modifications}
\begin{figure*}[t]
  \includegraphics[width=0.9\textwidth]{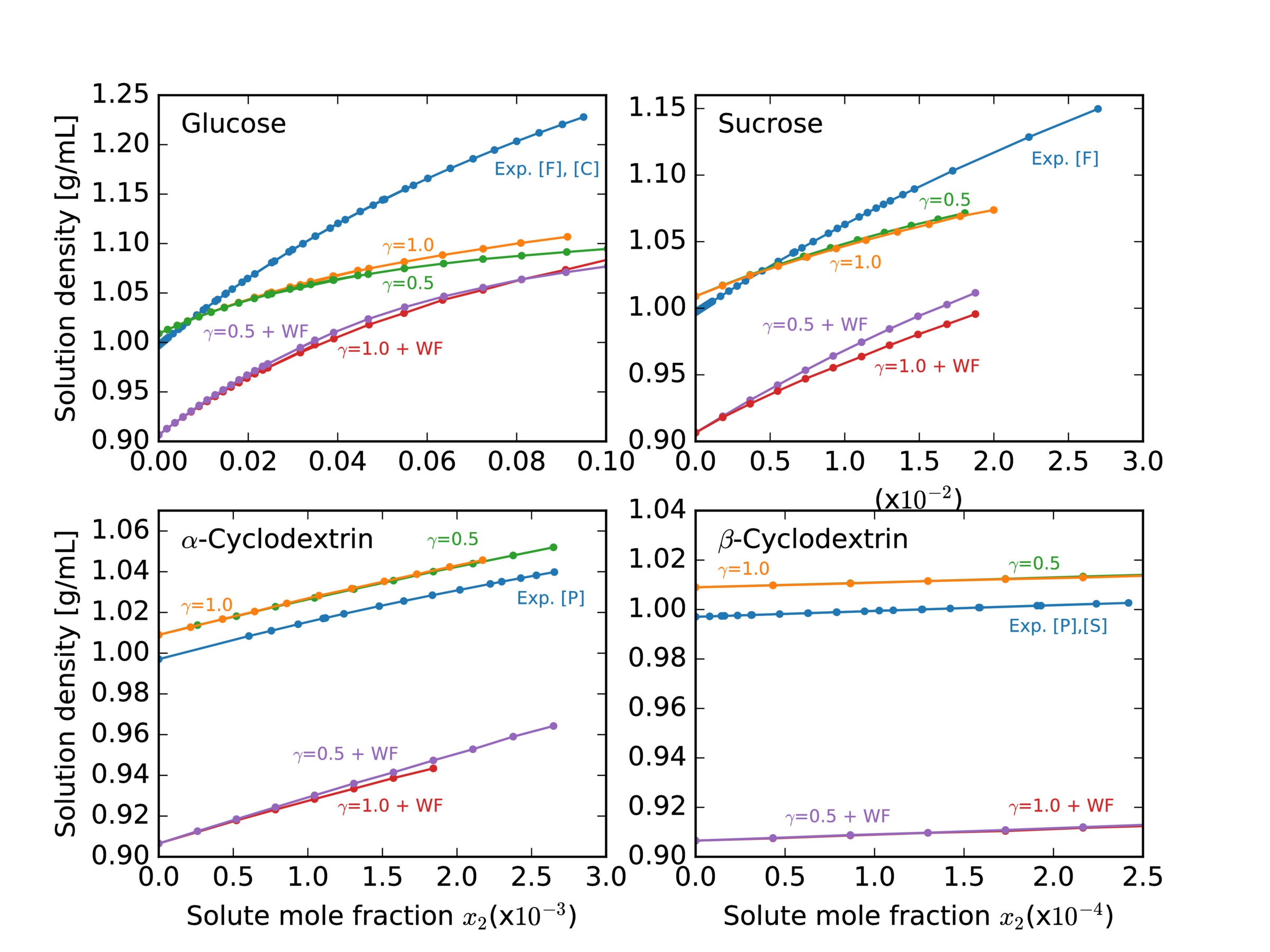}
  \caption{Dependence of solution density on solute mole fraction for different saccharides in the MARTINI model. Each panel shows experimental values (blue), the standard MARTINI model ($\gamma$=1.0) and the modified MARTINI model ($\gamma$=0.5) with pure standard MARTINI water (yellow and green curves respectively) and with antifreeze water (red and violet curves respectively). Note the different scaling of the x-axis in every panel. \changesfromto{none}{Experimental data were taken from references denoted [P]\cite{Paduano1990}, [C]\cite{Cerdeirina1997}, [F]\cite{Fucaloro2007}, and [S]\cite{Santos2016}.}}
  \label{fgr:density}
\end{figure*}

The density of a solution directly relates to the partial volume of the solute which, on a molecular scale, can be interpreted as the volume increase upon addition of a single solute molecule to a given solution. In a MD simulation, solution density depends on the interplay of solute-solute, solute-solvent and solvent-solvent non-bonded interactions. To test if modified solute-solute interactions compromised \changesfromto{solute-solvent or solvent-solvent interactions}{the non-bonded force balance}, we calculated the solution density for glucose, sucrose, $\alpha$- and $\beta$-CD in water over a range of concentrations and compared it with available experimental data. To this end, the system volume averaged over the last five nanoseconds of 15~ns trajectories was used to calculate the solution density which was plotted against the solute mole fraction (fig. \ref{fgr:density}). Comparison with experimental data\cite{Paduano1990,Cerdeirina1997,Fucaloro2007,Santos2016} shows that MARTINI tends to underestimate the density of saccharide solutions by up to 10\% at high solute concentrations (fig. \ref{fgr:density}A, B). This finding is in line with previous MARTINI simulations for aqueous glucose solutions\cite{Lopez2009}. At low concentrations (solute mole fraction $<$~0.4\%) this effect is overcompensated by a $\approx$ 1\% overestimation of the density of pure water in MARTINI (fig. \ref{fgr:density}C, D). The use of \changesfromto{10\% antifreeze beads}{antifreeze water} leads to an additional systematic underestimation of the density by about 10\%, as was stated before\cite{Marrink2007}. Importantly, the densities obtained from simulations with scaled $\epsilon_{ij}$ ($\gamma = 0.5$) closely followed the densities of the standard MARTINI model (deviations $<$~1.5\%)\changesfromto{. This indicates}{, indicating} that the proposed force field modification does not disturb the overall \changesfromto{balance between non-bonded solute-solute, solute-solvent, and solvent-solvent interactions.}{non-bonded force balance.}

\changesfromto{none}{We furthermore tested the effect of the proposed modification in organic solvents by studying systems of A2 glycan and glucose in the apolar solvent hexadecane. 
The expected low solubility of saccharides under such conditions\cite{Mazzobre2005} was captured not only by the original but also by the scaled MARTINI force field (fig.~\ref{SI:A2_hexadecane}), suggesting the validity of the proposed scaling also in organic solvents.}

\section{DISCUSSION}

\paragraph*{\changesfromto
    {Non-bonded interactions in CG force fields}
{MARTINI strongly overestimates the aggregation propensity of saccharides}}
Every force field is an approximation and therefore its application is limited to cases it has been designed for. For MARTINI this is certainly the simulation of biological macromolecules (proteins\cite{monticelli_martini_2008}, carbohydrates\cite{Lopez2009}, and DNA\cite{Uusitalo2015}). In contrast to AA force fields that are mainly being used (and have been parametrized) for simulations of individual macromolecules, the advantage of a CG force field such as MARTINI lies in the possibility of simulating ensembles of macromolecules on microsecond time scales. 
\changesfromto
    {The simulation of a lipid membrane patch containing several membrane protein molecules has recently been demonstrated\cite{Ingolfsson2016} promising future simulations at subcellular scale.}
{This is exemplified by the simulation of a lipid membrane patch containing several membrane protein molecules\cite{Ingolfsson2016} which holds promise for future simulations at subcellular scale.} An important aspect in these mesoscopic systems will be how macromolecules interact\changesfromto{: The}{ and thus the} accurate representation of non-bonded forces between macromolecules is of crucial importance. 

\changesfromto{none}
{We found that the MARTINI force field strongly overestimates the aggregation propensity of saccharides in aqueous solution. We analyzed this quantitatively for a single water model (antifreeze water) and qualitative tests suggested similar problems for the other two water models (standard water without antifreeze particles and polarizable water; fig.~\ref{SI:ions_WF}C, D).}
\changesfromto
    {We found that $B_{22}$ for the small molecule glucose, which is comparable in size to the reference molecules, is wrong in MARTINI (-1.2 versus +0.117~L/mol), suggesting that MARTINI overestimates the solute-solute interaction strength with regard to the solute-water interaction. Clearly, this systematic error can be additive and will be amplified in macromolecules that are ten to several hundred times larger than the reference molecules.}
{The deviations between experimental and predicted $B_{22}$ values rapidly increase with saccharide size, suggesting that an imbalance in the parametrization of the LJ interaction potentials of the MARTINI beads amplifies as the molecules grow bigger.} 
Hence it is perhaps not surprising that the propensity to aggregate of either proteins\cite{Stark2013} or saccharides, as shown in this study, is \changesfromto{dramatically}{strongly} overestimated in MARTINI.  
\changesfromto
    {Indeed, recent examples in the literature suggest that the aforementioned overestimation of saccharide-saccharide interaction leads to nonphysical behavior of saccharides.}
{As a consequence recent MARTINI simulations involving saccharides suggested a tendency to aggregate that most likely does not reflect physical reality:} \citeauthor{Kapla2016}\cite{Kapla2016} found that trehalose, a disaccharide similar to sucrose, forms clusters in MARTINI polarizable water. The small positive experimental $B_{22}$ of trehalose (table \ref{tbl:B22exp}), however, indicates net weak repulsion of trehalose molecules so that clustering at the reported concentrations (0.22~kg/kg water; solubility limit: 1.36~kg/kg water at 310~K\cite{Lammert1998}) seems very unlikely to represent true physical behavior.  
\citeauthor{LPS}\cite{LPS} simulated a model of the bacterial outer membrane in MARTINI containing lipopolysaccharide (LPS) molecules, whose polysaccharide components extend into the extracellular space. Similar to our observations with unscaled MARTINI, LPS molecules clustered and collapsed onto the membrane surface. Again, this disagrees with experiments describing LPS as a polymer brush extending tens to a few hundred nanometers into the extracellular space\cite{Ivanov2011}, suggesting instead an overestimation of \changesfromto{saccharide-saccharide, and possibly saccharide-lipid, attraction.}{the aggregation propensity of saccharides and/or lipids.}      

\paragraph*{\changesfromto
    {How non-bonded forces have been parametrized in MARTINI}
{Validity of the proposed correction, other approaches}}
\changesfromto
   {The parametrization of the non-bonded van der Waals forces in MARTINI is largely based on solvation free energies of small organic reference molecules, {\it i.e.} solute-solvent interactions\cite{Marrink2007}, with particular regard to the accurate reproduction of partition coefficients of these molecules in diverse water/organic solvent systems.
   The tendency of a solute to aggregate or disperse in water, represented by $B_{22}$, depends on the ratio of solute-water to solute-solute interactions; both are known to be systematically too low for the reference molecules in MARTINI\cite{Marrink2007}.}
{The proposed scaling of the solute-solute interaction is an {\it{ad hoc}} remedy to correct the imbalance of non-bonded forces in the MARTINI force field in combination with a particular water model (antifreeze water). It suffices to abrogate unrealistic aggregation behavior of saccharides in aqueous solution, retains the original solvent partitioning behavior, and is compatible with earlier findings for proteins\cite{Stark2013}. 
The underlying cause for the non-bonded force imbalance, however, cannot be limited to the strength of solute-solute interactions alone as is evident by the failure to quantitatively reproduce $B_{22}$ values for larger saccharides. Two extreme tests illustrate the general limit of scaling solute-solute interactions: Neither doubling $\sigma_{solute-solute}$ (which, given the cutoff value and the potential shift, reduces $\epsilon_{solute-solute}$ to almost zero), nor the conversion of all beads in A2 glycan to special S beads (scaling $\epsilon_{solute-solute}$ to 75\% and $\sigma_{solute-solute}$ to 0.43 nm, but violating the 4:1 mapping rule in MARTINI\cite{Marrink2007}), lifted $B_{22}$ beyond 15~L/mol (figs.~\ref{SI:B22_sigma} and \ref{SI:A2_S-beads}). 
These findings are indicative of an existing imbalance between solute-water and water-water interactions. It is known from calculations of hydration free energies that the ratio of solute-water/water-water interaction strengths is too low with MARTINI standard water\cite{Marrink2007}. The ratio is even lower with antifreeze water whose average water-water interaction strength is slightly higher. It is not enough though to simply change the water model to standard water or polarizable water as shown by the rapid formation of saccharide aggregates also in these systems, albeit the magnitude of the imbalance is smaller (fig.~\ref{SI:ions_WF}C, D). Moreover, the water model in MARTINI is known to underestimate the water-water interaction strength in the liquid phase\cite{Marrink2007}.
Hence, both solute-water and solute-solute interactions need to be reparametrized, possibly in combination with an improved water model, to reach experimental $B_{22}$ values and thus more realistic molecular interaction behavior. Also, we recommend to include not only single beads as chemical building blocks in the parametrization, but biological reference macromolecules (proteins, DNA, saccharides) to avoid an amplification of small errors as has become evident in this study.}

\paragraph*{Non-bonded interactions in AA force fields and conclusion}
CG simulations, including MARTINI, are often verified based on agreement of selected observables with an atomistic approach and the same could be done, in principle, for solute-solute interactions. \changesfromto
   {Recently it has been shown, however, that bio-macromolecule aggregation and compaction in practically all modern AA force fields is also troubled with strong overestimation of inter-protein interaction strengths}
{Recently, however, overestimation of protein aggregation propensity has been reported in practically all modern AA force fields}
\cite{Best_2014_aggregation,Jao_2015_aggregation,Petrov2014,Abriata2015,Miller2016_aa}. It has been argued that this is at least partially due to the way AA force fields are fine-tuned, {\it i.e.} to maintain protein native structure, leading to too strong protein-protein non-bonded interactions, and suggest scaling of solute-solute/solute-solvent interactions or partial charges adjustment as a remedy. \changesfromto
    {none}
{In many of these cases this nonphysical behavior could also stem from the use of certain water models (SPC, TIP3P), which have been shown\cite{Best_2014_aggregation,Miller2016_aa} to promote solute aggregation.}
Parenthetically, atomistic simulations of A2 glycan molecules with the GLYCAM06j force field and SPC water (for details see Supplementary Information and consult fig.~\ref{SI:snap_A2_AA_CG1.0_0.5}) resulted in aggregating behavior very similar to unscaled MARTINI simulations, suggesting similar imperfections for saccharides and certain water models. 

Indeed, the Grafm{\"u}ller group found nonphysical aggregation of monosaccharides (glucose, mannose, xylose) \cite{Sauter2015} for GLYCAM06h with TIP3P water\cite{TIP3P} which could largely be corrected by use of the TIP5P water model\cite{TIP5P}. The Elcock group\cite{Lay2016_carb} confirmed the finding for GLYCAM06/TIP3P with glucose and sucrose and found a very similar deficit for CHARMM36-TIP3P. They achieved correction by a substantial reduction of $\epsilon$ in LJ interactions between C-O and C-C atoms in the GLYCAM06 force field. It will be interesting to see if the newly developed TIP4P-D water\cite{Piana2015}, demonstrated to improve the compaction of disordered proteins, will also aid saccharide simulations.

\changesfromto
    {We conclude that any force field intended for MD simulations of ensembles of biological macromolecules needs be (re-)parametrized taking into account macroscopic observables of macromolecular systems, like $B_{22}$, solubility or even mechanical properties of molecules.}
{We conclude that scaling saccharide-saccharide interactions provides an {\it{ad hoc}} solution to remedy nonphysical aggregation behavior in MARTINI 2.x. Due to the existing non-bonded force imbalance a reparametrization of the entire force field, with particular regard to the water model, will be necessary to facilitate quantitative predictions of aggregation propensities.}
Otherwise, bold simulation attempts of systems at micron scale\cite{Feig2015,Feig2016} run the risk of producing drastically misleading results. \changesfromto{none}{Furthermore, macroscopic observables of macromolecular systems, like $B_{22}$, solubility or even mechanical properties of molecules, need to be taken into account in the reparametrization of MARTINI to ensure its suitability for MD simulations of ensembles of macromolecules.}\\

\section*{Acknowledgment}
The authors thank Antje Potthast, Marek Cieplak, Tomasz W{\l}odarski, and Damien Thompson for fruitful discussions and the IST Austria Scientific Computing Facility for support. P.S.S. was supported by research fellowship 2811/1-1 from the German Research Foundation (DFG) and M.S. was supported by EMBO Long Term Fellowship ALTF 187-2013 and a grant no GC65-32 from the Interdisciplinary Centre for Mathematical and Computational Modelling (ICM), University of Warsaw, Poland.
\bibliography{ref}

\newpage

\pagebreak
\widetext
\begin{center}
\textbf{\large Supplemental Materials: Overcoming the limitations of the MARTINI force field in Molecular Dynamics simulations of polysaccharides}
\end{center}
\setcounter{figure}{0}
\setcounter{table}{0}
\makeatletter
\renewcommand{\theequation}{S\arabic{equation}}
\renewcommand{\thefigure}{S\arabic{figure}}
\renewcommand{\thetable}{S\arabic{table}}
\renewcommand{\theHtable}{Supplement.\thetable}
\renewcommand{\theHfigure}{Supplement.\thefigure}

Supporting tables S1 and S2 and supporting figures S1-S5 are provided below. 
The ancillary files \textbf{A2\_bonds\_AA\_vs\_MARTINI.pdf}, \textbf{A2\_angles\_AA\_vs\_MARTINI.pdf}, and \textbf{A2\_dihedrals\_AA\_vs\_MARTINI.pdf} contain distributions of bond lengths, angles, and dihedrals, respectively, for an {\it{a posteriori}} coarse-grained AA simulation of A2 glycan (150 ns, black traces) and a MARTINI simulation without dihedral potentials switched on (red traces) for all bonded interactions. The \textbf{A2\_dihedrals\_AA\_vs\_MARTINI.pdf} file contains in addition dihedral distributions for a MARTINI simulation with dihedral potentials (gray traces). Similarly, the files  \textbf{ACD\_bonds\_AA\_vs\_MARTINI.pdf},  \textbf{ACD\_angles\_CG\_vs\_MARTINI.pdf} and  \textbf{BCD\_bonds\_AA\_vs\_MARTINI.pdf},  \textbf{BCD\_angles\_AA\_vs\_MARTINI.pdf} contain AA and MARTINI distributions of unique bonds and angles for $\alpha$-CD and $\beta$-CD systems respectively. \\   

\section*{Supplementary Tables}

\begin{table}[!ht]
\tiny
  \caption{MARTINI bead types for coarse-grained carbohydrates}
  \label{tbl:bead_types}
  \begin{tabular}{llllllllll}
    \hline
    saccharide             & bead no.$^a$  & bead name$^{a,b}$   & bead type$^c$\\
    \hline
    $\beta$-D-glucose      & 1             & B3              & GP4    \\
                           & 2             & B2              & GP4    \\
                           & 3             & B1              & GP1    \\
    \hline
    sucrose                & 1             & B1              & GP1        \\
                           & 2             & B2              & GP2        \\
                           & 3             & B3              & GP4        \\
                           & 4             & B4              & GP1        \\
                           & 5             & B5              & GP1        \\
                           & 6             & B6              & GP4        \\
    \hline
    $\alpha$-cyclodextrin  & 1             & B1              & GP1        \\
                           & 2             & B2              & GP2        \\
                           & 3             & B3              & GP4        \\
    \hline
    $\beta$-cyclodextrin   & 1             & B1              & GP1        \\
                           & 2             & B2              & GP2        \\
                           & 3             & B3              & GP4        \\
    \hline
    A2 glycan              & 1             & Gn11            & GNa        \\
                           & 2             & Gn12            & GP4        \\
                           & 3             & Gn13            & GSP1       \\
                           & 4             & Gn14            & GSP1       \\
                           & 5             & Gn21            & GP5        \\
                           & 6             & Gn22            & GSP1       \\
                           & 7             & Gn23            & GP1        \\
                           & 8             & bM1             & GP1        \\
                           & 9             & bM2             & GSP1       \\
                           & 10            & bM3             & GNda       \\
                           & 11            & aM31            & GNda       \\
                           & 12            & aM32            & GP4        \\
                           & 13            & aM33            & GSP1       \\
                           & 14            & aM61            & GNda       \\
                           & 15            & aM62            & GP4        \\
                           & 16            & aM63            & GSP1       \\
                           & 17            & Gn31            & GP5        \\
                           & 18            & Gn32            & GNda       \\
                           & 19            & Gn33            & GSP1       \\
                           & 20            & Ga31            & GSP1       \\                       
                           & 21            & Ga32            & GSP1       \\
                           & 22            & Ga33            & GP1        \\
                           & 23            & SA31            & GQa        \\
                           & 24            & SA32            & GSP1       \\
                           & 25            & SA33            & GSP1       \\
                           & 26            & SA34            & GP5        \\
                           & 27            & SA35            & GP4        \\
                           & 28            & Gn61            & GP5        \\
                           & 29            & Gn62            & GNda       \\
                           & 30            & Gn63            & GSP1       \\                       
                           & 31            & Ga61            & GSP1       \\
                           & 32            & Ga62            & GSP1       \\
                           & 33            & Ga63            & GP1        \\
                           & 34            & SA61            & GQa        \\
                           & 35            & SA62            & GSP1       \\
                           & 36            & SA63            & GSP1       \\
                           & 37            & SA64            & GP5        \\
                           & 38            & SA65            & GP4        \\
                          
    \hline
  \end{tabular}
\end{table}
\footnotesize{
$^a$cf. fig. \ref{fgr:mapping} in the main text\\
$^b$bead names prefixed with "B" follow \citeauthor{Lopez2009}\\ 
$^c$bead types are according to \citeauthor{Marrink2007}; prefix "G" means LJ interaction with other "G" beads is scaled according to \eqref{eqn:gamma_def}. See main text for details.}

\cleardoublepage
\begin{table}
\begin{minipage}{\textwidth}                          
\tiny
  \caption{Force Field Parameters for coarse-grained carbohydrates$^a$
  }  
  \label{tbl:params}
  \begin{tabular}{llllllllll}  
    \hline
    saccharide              & bond      & $r_{eq}$& $k_{r}$ & angle & $\theta_{eq}$ & $k_\theta$ & dihedral & $\phi_{eq}$ & $k_{\phi}$\\
                            &           & [nm]  & [$\frac{kJ}{mol\cdot nm^2}$] & & [deg] & [$\frac{kJ}{mol\cdot[rad]^2}$] & & [deg] & [$\frac{kJ}{mol\cdot[rad]^2}$] \\
    \hline  
    $\beta$-d-glucose       & B3-B2     & 0.323 & $\infty$ & & & & & &\\
                            & B3-B1     & 0.384 & $\infty$ & & & & & &\\
                            & B2-B1     & 0.331 & $\infty$ & & & & & &\\
\hline
    sucrose                 & B1-B2     & 0.222 & $\infty$ & B1-B2-B4       & 130 & 10 & B1-B2-B4-B5 & 130 & 25\\
                            & B2-B3     & 0.247 & $\infty$ & B3-B2-B4       & 110 & 150& B1-B2-B4-B6 & 80  & 2 \\
                            & B2-B4     & 0.429 & $\infty$ & B5-B4-B2       & 20  & 50 & B3-B2-B4-B5 & -70 & 20\\
                            & B4-B5     & 0.293 & $\infty$ & B6-B4-B2       & 85  & 150& & &\\
                            & B4-B6     & 0.372 & $\infty$ &                &     &    & & &\\
\hline
    $\alpha$-cyclodextrin   & G11-G12   & 0.215 & 5500     & G11-G12-G22    & 80  &  1 & & &\\
                            & G12-G13   & 0.220 & 2500     & G12-G22-G32    & 120 & 28 & & &\\
                            & G12-G22   & 0.460 & 450      & G23-G22-G12    & 94  & 100& & &\\
\hline                            
    $\beta$-cyclodextrin    & G11-G12   & 0.215 & 10000    & G11-G12-G22    & 70  & 5  & & &\\
                            & G12-G13   & 0.220 & $\infty$ & G12-G22-G32    & 135 & 30 & & &\\
                            & G12-G22   & 0.470 & 500      & G23-G22-G12    & 90  & 120& & &\\
\hline
    A2 glycan               & Gn12-Gn13 & 0.270 & 8000     & Gn11-Gn12-Gn13 & 120  & 250& Gn23-Gn21-Gn13-Gn12& 180&20\\
                            & Gn13-Gn14 & 0.340 & 12000    & Gn11-Gn12-Gn14 & 147 & 500& bM3-bM1-Gn22-Gn21 & 80&10\\
                            & Gn12-Gn14 & 0.380 & 20000    & Gn12-Gn13-Gn21 & 154 & 400& aM33-aM31-bM2-bM1 & -10&40\\
                            & Gn13-Gn21 & 0.433 & 5000     & Gn14-Gn13-Gn21 & 72  & 250& aM63-aM61-bM3-bM1 & 180&40\\
                            & Gn21-Gn22 & 0.378 & 10000    & Gn13-Gn21-Gn22 & 86  & 600& Gn33-Gn31-aM31-aM33& 170&40\\
                            & Gn21-Gn23 & 0.524 & 22000    & Gn13-Gn21-Gn23 & 50  & 300& Ga33-Ga31-Gn32-Gn31& -160&10\\
                            & Gn22-bM1  & 0.338 & 15000    & Gn21-Gn22-bM1  & 168 & 500& SA33-SA31-Ga32-Ga31& -50&5\\
                            & bM1-bM3   & 0.334 & 18000    & Gn23-Gn22-bM1  & 90  & 50 & Gn63-Gn61-aM61-aM63& 160&35\\
                            & bM2-aM31  & 0.372 & 9000     & Gn22-bM1-bM2   & 131 & 700& Ga63-Ga61-Gn62-Gn61& 110&8\\
                            & aM32-aM33 & 0.322 & 16000    & Gn22-bM1-bM3   & 78  & 200& SA63-SA61-Ga62-Ga61& 20&30\\
                            & aM31-aM33 & 0.349 & 8500     & bM1-bM2-aM31   & 105 & 200& & &\\
                            & bM3-aM61  & 0.354 & 8000     & bM3-bM2-aM31   & 169 & 600& & &\\
                            & aM62-aM63 & 0.323 & 16000    & bM2-aM31-aM32  & 110 & 300& & &\\
                            & aM61-aM63 & 0.354 & 15000    & bM2-aM31-aM33  & 86  & 200& & &\\
                            & aM31-Gn31 & 0.340 & 4500     & bM2-aM31-Gn31  & 122 & 60 & & &\\
                            & Gn31-Gn32 & 0.388 & 20000    & aM33-aM31-Gn31 & 90  & 80 & & &\\
                            & Gn32-Gn33 & 0.315 & 20000    & aM31-Gn31-Gn32 & 97  & 550& & &\\
                            & Gn31-Gn33 & 0.516 & 15000    & aM31-Gn31-Gn33 & 61  & 400& & &\\
                            & Ga32-Ga33 & 0.312 & 12000    & Gn31-Gn32-Ga31 & 167 & 700& & &\\
                            & Ga32-SA31 & 0.335 & 2000     & Gn33-Gn32-Ga31 & 78  & 250& & &\\
                            & SA31-SA33 & 0.382 & 15000    & Gn32-Ga31-Ga32 & 112 & 500& & &\\
                            & SA32-SA34 & 0.357 & 15000    & Gn32-Ga31-Ga33 & 71  & 350& & &\\
                            & SA33-SA35 & 0.296 & 10000    & Ga31-Ga32-SA31 & 108 & 180& & &\\
                            & aM61-Gn61 & 0.340 & 4800     & Ga33-Ga32-SA31 & 137 & 50 & & &\\
                            & Gn61-Gn62 & 0.388 & 20000    & Ga32-SA31-SA32 & 100 & 250& & &\\
                            & Gn62-Gn63 & 0.315 & 20000    & Ga32-SA31-SA33 & 80  & 220& & &\\
                            & Gn61-Gn63 & 0.516 & 15000    & SA31-SA32-SA34 & 122 & 800& & &\\
                            & Ga62-Ga63 & 0.320 & 12000    & SA31-SA33-SA35 & 109 & 350& & &\\
                            & Ga62-SA61 & 0.335 & 2000     & SA32-SA33-SA35 & 157 & 400& & &\\
                            & SA61-SA63 & 0.382 & 15000    & SA33-SA32-SA34 & 70  & 80 & & &\\
                            & SA62-SA64 & 0.357 & 15000    & bM1-bM3-aM61   & 90  & 150& & &\\
                            & SA63-SA65 & 0.294 & 10000    & bM2-bM3-aM61   & 117 & 150& & &\\
                            & Gn11-Gn12 & 0.280 & 4000     & bM3-aM61-aM62  & 118 & 400& & &\\
                            & Gn22-Gn23 & 0.312 & $\infty$ & bM3-aM61-aM63  & 83  & 50 & & &\\
                            & bM1-bM2   & 0.276 & $\infty$ & bM3-aM61-Gn61  & 119 & 190& & &\\
                            & bM2-bM3   & 0.307 & $\infty$ & aM61-Gn61-Gn62 & 80  & 100& & &\\
                            & aM31-aM32 & 0.277 & $\infty$ & aM61-Gn61-Gn63 & 61  & 600& & &\\
                            & aM61-aM62 & 0.277 & $\infty$ & Gn61-Gn62-Ga61 & 140 & 50 & & &\\
                            & Gn32-Ga31 & 0.365 & $\infty$ & Gn63-Gn62-Ga61 & 78  & 250& & &\\
                            & Ga31-Ga32 & 0.269 & $\infty$ & Gn62-Ga61-Ga62 & 112 & 600& & &\\
                            & Ga31-Ga33 & 0.399 & $\infty$ & Gn62-Ga61-Ga63 & 71  & 300& & &\\
                            & SA31-SA32 & 0.337 & $\infty$ & Ga61-Ga62-SA61 & 107 & 230& & &\\
                            & SA32-SA33 & 0.315 & $\infty$ & Ga63-Ga62-SA61 & 135 & 200& & &\\
                            & Gn62-Ga61 & 0.366 & $\infty$ & Ga62-SA61-SA62 & 99  & 170& & &\\
                            & Ga61-Ga62 & 0.269 & $\infty$ & Ga62-SA61-SA63 & 80  & 170& & &\\
                            & Ga61-Ga63 & 0.399 & $\infty$ & SA61-SA62-SA64 & 122 & 700& & &\\
                            & SA61-SA62 & 0.336 & $\infty$ & SA61-SA63-SA65 & 109 & 200& & &\\
                            & SA62-SA63 & 0.315 & $\infty$ & SA62-SA63-SA65 & 157 & 350& & &\\
                            &           &       &          & SA63-SA62-SA64 & 70  & 80 & & &\\
excluded non-bonded         & Gn13-Gn23 &       &          &                &     &    & & &\\
interactions                & Gn14-Gn21 &       &          &                &     &    & & &\\
                            & Gn22-bM3  &       &          &                &     &    & & &\\
                            & Gn23-bM1  &       &          &                &     &    & & &\\
                            & bM1-aM61  &       &          &                &     &    & & &\\
                            & bM3-aM63  &       &          &                &     &    & & &\\
                            & aM31-Gn33 &       &          &                &     &    & & &\\
                            & aM33-Gn31 &       &          &                &     &    & & &\\
                            & aM61-Gn63 &       &          &                &     &    & & &\\
                            & Gn32-Ga33 &       &          &                &     &    & & &\\
                            & SA33-SA34 &       &          &                &     &    & & &\\
                            & Gn62-Ga63 &       &          &                &     &    & & &\\
                            & SA63-SA64 &       &          &                &     &    & & &\\              
    \hline
  \end{tabular}
\end{minipage}
\end{table}
\footnotesize{$^a$In the Table~\ref{tbl:params} bond spring constants equal $\infty$ denote that the bond was turned into a constraint. Non-bonded interactions were excluded between selected beads to achieve good agreement of some of the angle distributions with AA simulations.}

\newpage

\section*{Supplementary Figures}

\begin{figure}
\includegraphics[width=0.9\textwidth]{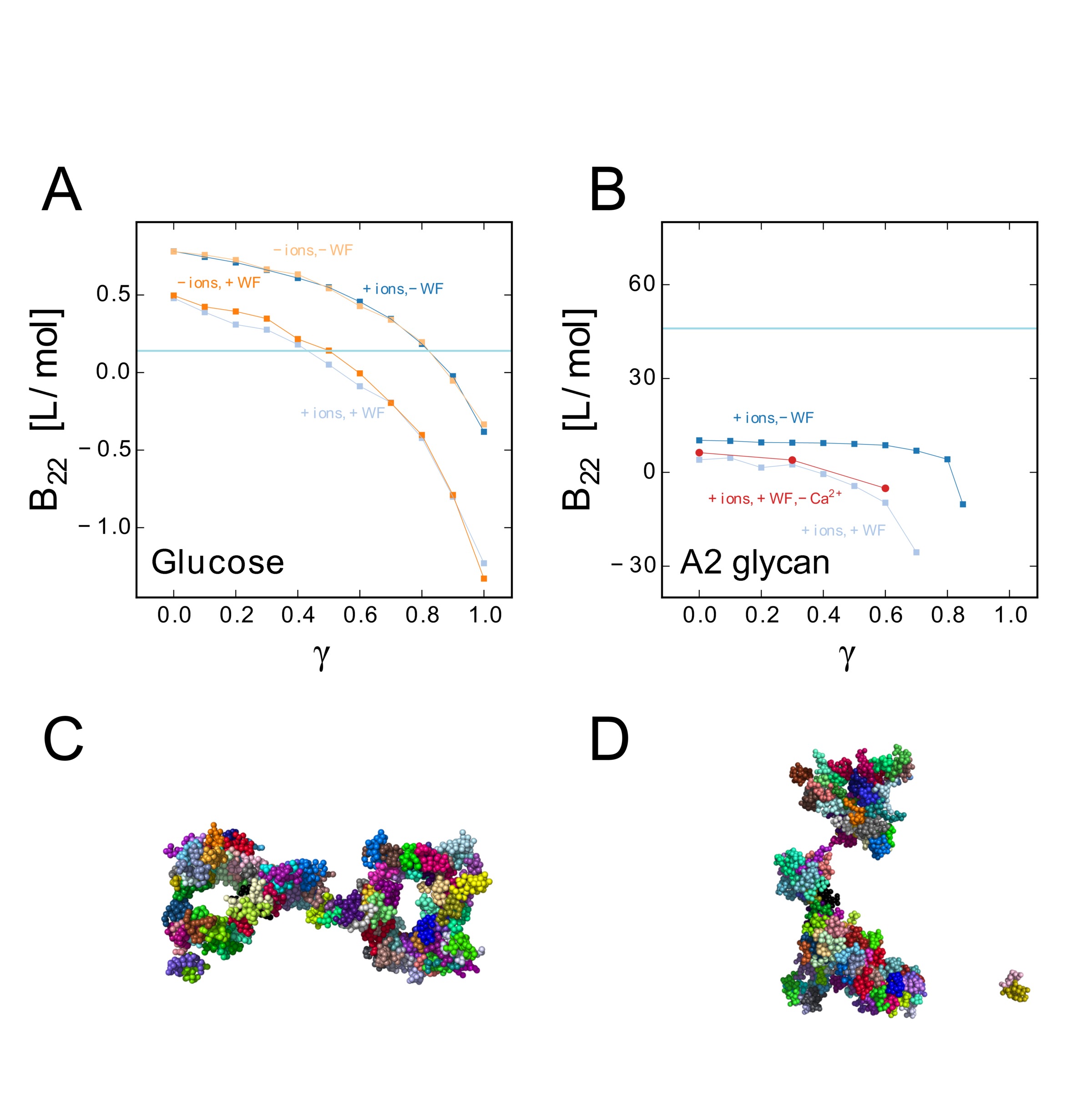}
\caption{The $B_{22}(\gamma )$ dependence on the content of the solvent for glucose (A) and A2 glycan (B). Note, that $B_{22}$ values are provided only for conditions where saccharides remain soluble. Different curves illustrate different solvation conditions, as described in the plot. In particular, ion content does have a negligible influence on $B_{22}$, independently on whether bivalent ions (Ca$^{2+}$) are present or not. Omission of the antifreeze beads (WF), however, increases $B_{22}$ by weakening water-water interactions, yet the qualitative behaviour of A2 glycan remains unaltered, as seen in the snapshot (C). A2 glycan is the only saccharide studied here endowed with charges, therefore its behaviour should depend on dielectric properties of the solvent. Introduction of polarizable water, however, had similarly negligible effect and prompt aggregation still occurred (D). }
\label{SI:ions_WF}
\end{figure}

\begin{figure}
\includegraphics[width=0.9\textwidth]{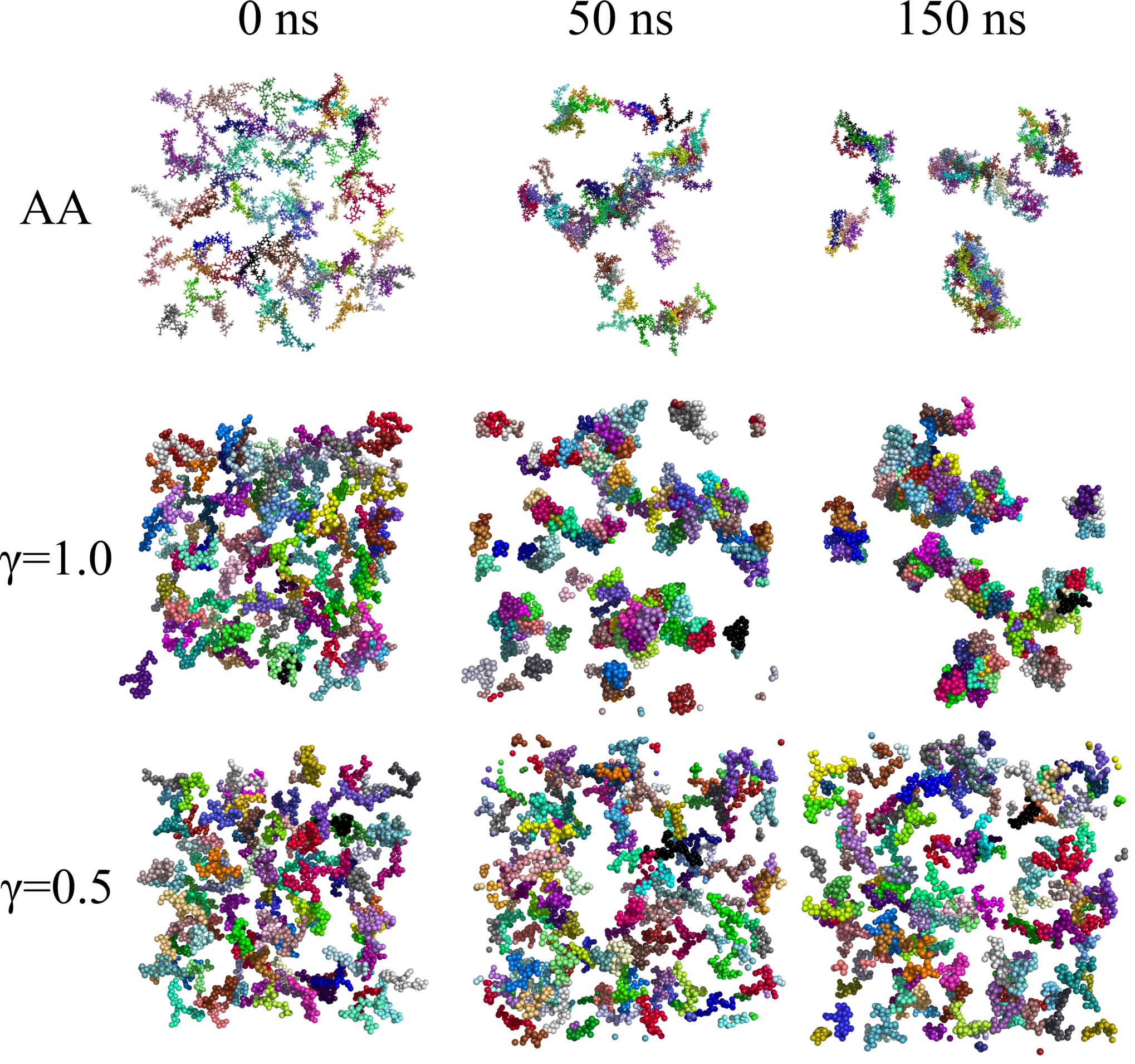}
\caption{Snapshots from AA and unscaled ($\gamma=1.0$) and scaled ($\gamma=0.5$) MARTINI simulations for 105 A2 glycan molecules in water. Each A2 glycan molecule is represented in a different color. Simulation time is given above each column. Note that the cut-off value for both LJ and Coulomb interactions was adjusted to 0.9~nm.}
\label{SI:snap_A2_AA_CG1.0_0.5}
\end{figure}

\begin{figure}
\includegraphics[width=0.9\textwidth]{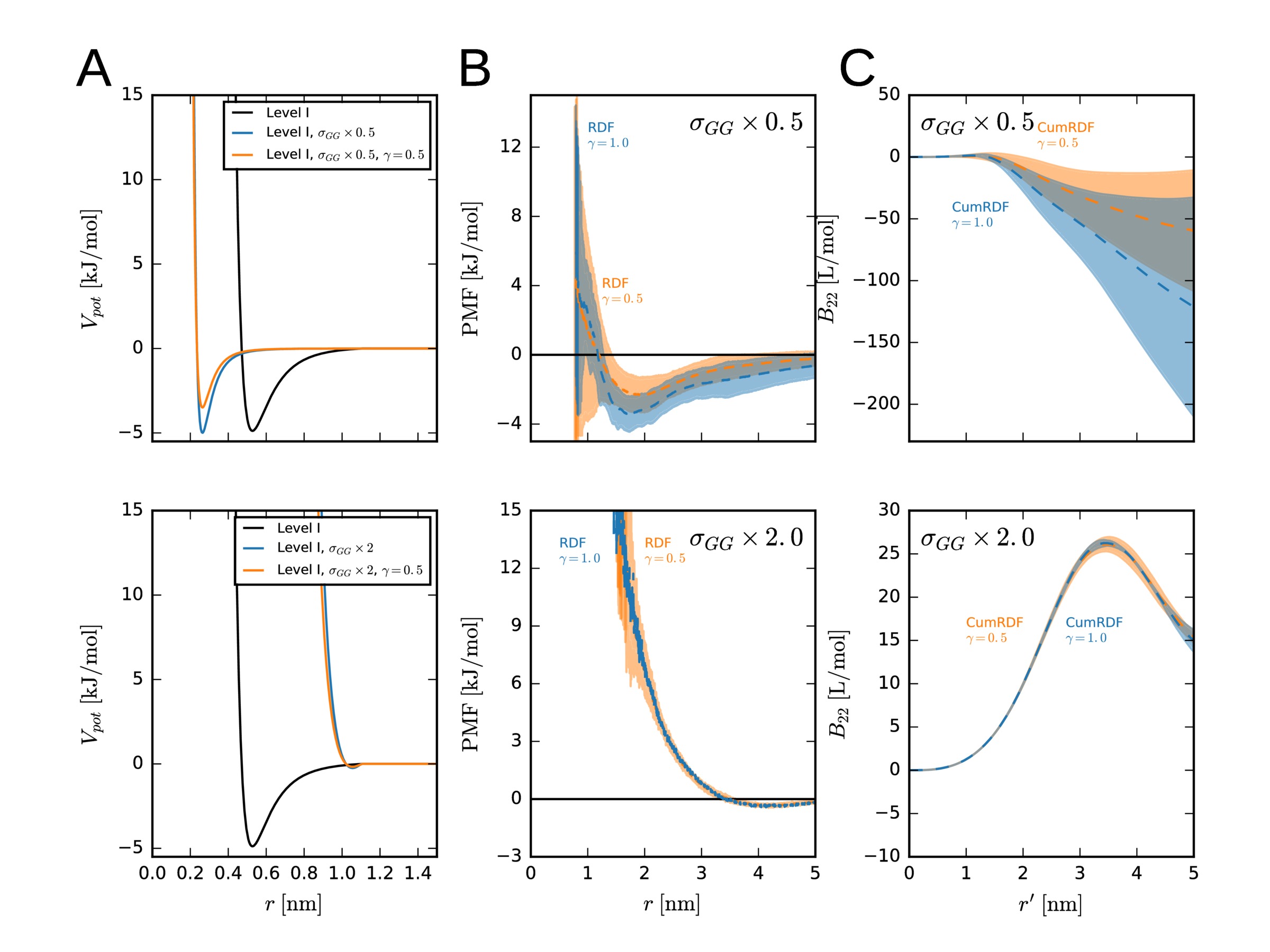}
\caption{Dependence of $B_{22}$ of A2 glycan on $\sigma_{GG}$. A, LJ potential of a typical saccharide-saccharide bead interaction (MARTINI level I, $\sigma=0.47$ nm, $\epsilon=5$ kJ/mol\cite{Marrink2007}, cutoff=1.1~nm, potential shifted to zero at cutoff) scaled with a factor of either 0.5 (top) or 2 (bottom). B, PMF as calculated from RDF (105 molecules, (19 nm)$^3$), WF, 3.1 $\mu$s) as described in the main text. C, $B_{22}$ calculated from the cumulative RDF of the same simulation as described in the main text. The RDF simulations with the reduced $\sigma$ did not fully converge within the simulated time and thus should be treated with caution. Transient aggregates observed during visual inspection point to even lower values of $B_{22}$. }
\label{SI:B22_sigma}
\end{figure}

\begin{figure}
\includegraphics[width=0.9\textwidth]{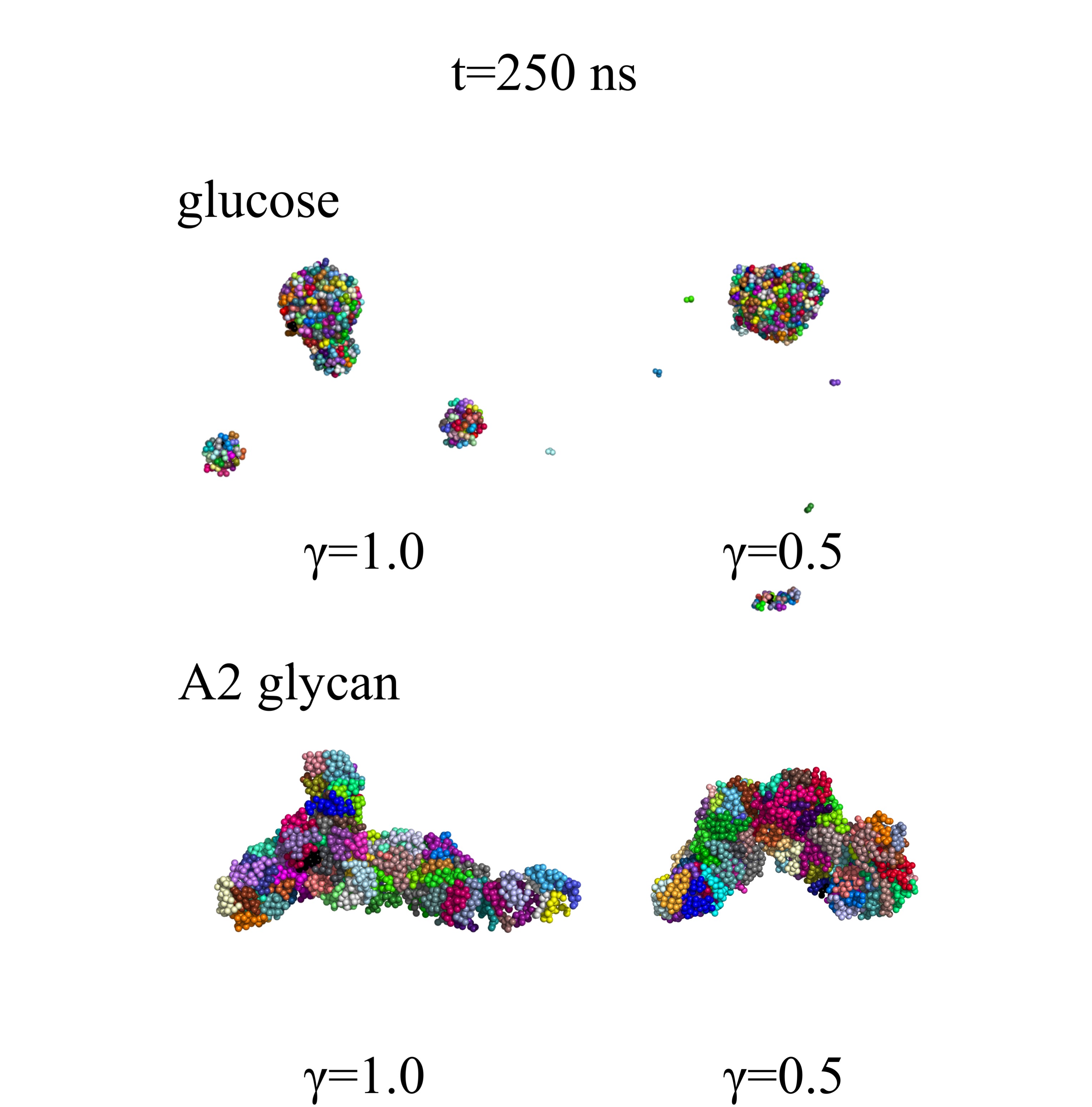}
\caption{Snapshots after 250 ns of MARTINI simulations of glucose and A2 glycan in the apolar solvent hexadecane. Note that counterions (Na$^+$) were added to balance the negative charge of A2 glycan. In all cases saccharides readily aggregate, following the expected behavior in apolar solvents\cite{Mazzobre2005}.}
\label{SI:A2_hexadecane}
\end{figure}

\begin{figure}
\includegraphics[width=0.9\textwidth]{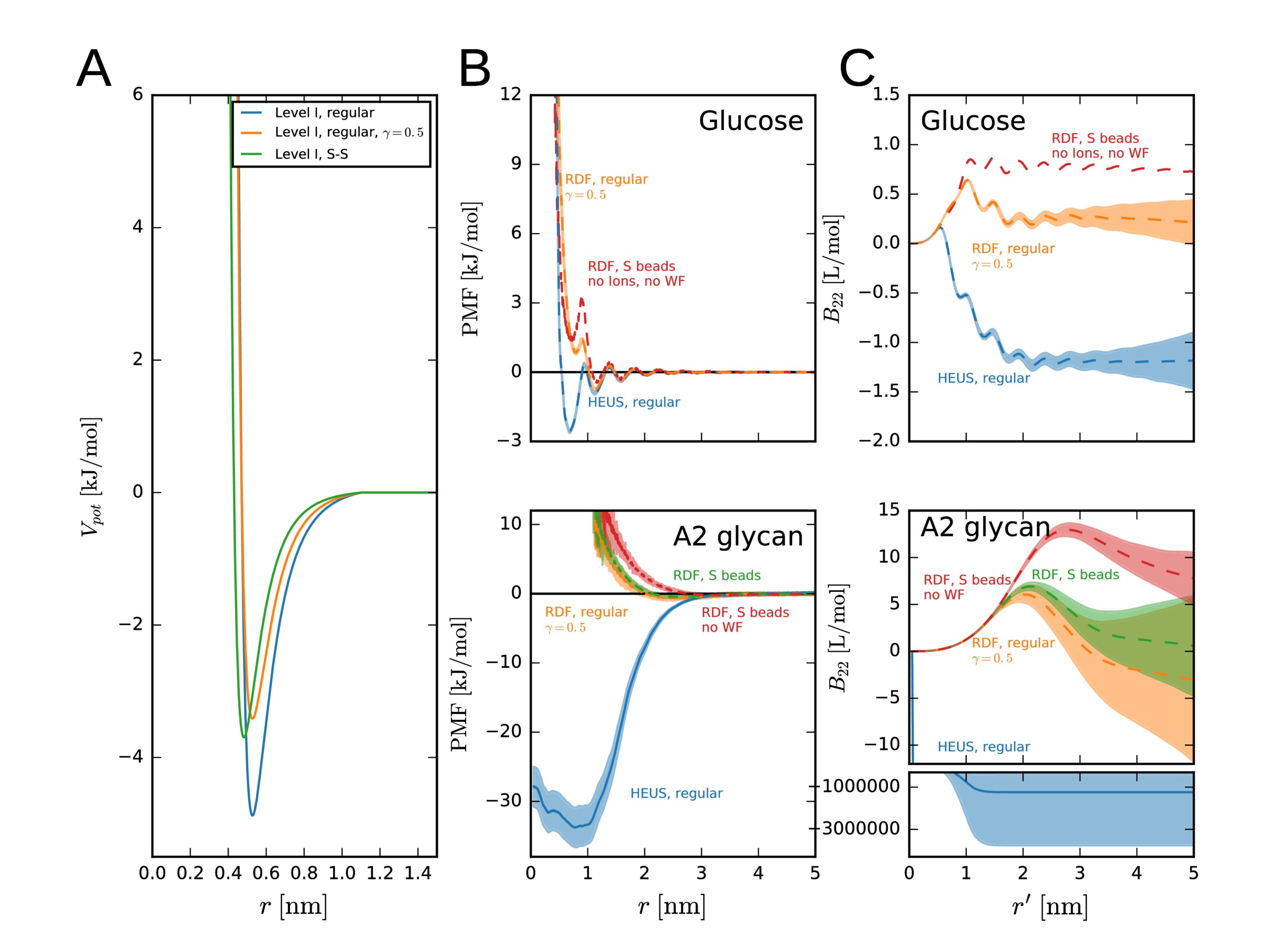}
\caption{MARTINI simulations of glucose and A2 glycan parametrized entirely with ring particles (S beads) endowed with 75\% of the LJ interaction strength ($\epsilon$) of their regular counterparts towards other S beads and $\sigma=0.43$ nm, as opposed to the regular $\sigma=0.47$ nm. A, LJ potential of a typical saccharide-saccharide bead interaction (MARTINI level I, $\sigma=0.47$ nm, $\epsilon=5$ kJ/mol; blue), the same interaction scaled with $\gamma$=0.5 (see main text, orange), and the corresponding level I interaction between two S beads ($\sigma=0.43$ nm, $\epsilon=3.5$ kJ/mol; green). All potentials are shifted to zero at cutoff=1.1~nm. B, PMF for glucose (top panel) and A2 glycan (bottom panel) with every bead converted to an S bead, calculated from the RDF (420 or 105 molecules, (19 nm)$^3$), $>$ 1 $\mu$s) from simulations with (green) or without (red) antifreeze particles as described in the main text. For comparison the unscaled and scaled PMFs from main fig. \ref{fgr:B22} are shown in blue and orange, respectively. C, Corresponding $B_{22}$ calculated from the RDFs in panel B as described in the main text.}
\label{SI:A2_S-beads}
\end{figure}
\end{document}